\documentclass[final, journal]{IEEEtran}
\usepackage{makeidx}  % allows for index generation
\usepackage{color,xcolor,soul,colortbl}
\usepackage[]{hyperref} % enables typesetting of hyperlinks: hidelinks, draft
\usepackage[all]{hypcap} % fix a problem of hyperref when link to a float

\usepackage{bm}
\usepackage{hyphenat}

\usepackage{natbib}	
\usepackage{booktabs}
\usepackage{tikz}  % use for drawing arrows

\usepackage{fancyhdr}
\usepackage{mathtools}
\usepackage{amsmath}

\usepackage{float} 			% used for figures and tables
\usepackage{subcaption}
\usepackage{graphicx,psfrag,epsfig}
\graphicspath{{./figures/}} % address of figures
\usepackage{sidecap}
\usepackage{epstopdf}
\usepackage[labelfont=bf]{caption}
\usepackage{array}
\usepackage{multirow} % use multirow inside a table
\usepackage{makecell} % force to break a cell

\newcolumntype{L}[1]{>{\raggedright\let\newline\\\arraybackslash\hspace{0pt}}m{#1}}
\newcolumntype{C}[1]{>{\centering\let\newline\\\arraybackslash\hspace{0pt}}m{#1}}
\newcolumntype{R}[1]{>{\raggedleft\let\newline\\\arraybackslash\hspace{0pt}}m{#1}}

\usepackage[author={Hessam}]{pdfcomment} % this package is used to add some comments on the pdf.

\usepackage{xargs}  % Use more than one optional parameter in a new commands
\usepackage{xparse}  % So that we can have two optional parameters
\usepackage{textcmds} % provides shorthand commands

\definecolor{LightCyan}{rgb}{0.88,1,1}
\definecolor{LightGreen}{rgb}{0.88,1,0.88}
\definecolor{DarkGreen}{rgb}{0.0, 0.7, 0.1}
\definecolor{CommentColor}{rgb}{0.9, 0.3, 0.7}

\newcommand{\tcb}[1]{\textcolor{DarkGreen}{#1}}

\newcommand*{\rom}[1]{\expandafter\@slowromancap\romannumeral #1@}
\newcommand{\RNum}[1]{\uppercase\expandafter{\romannumeral #1\relax}}

\NewDocumentCommand\DownArrow{O{2.0ex} O{black}}{%
   \mathrel{\tikz[baseline] \draw [<-, line width=0.5pt, #2] (0,0) -- ++(0,#1);}
}
\hypersetup{
    colorlinks,
	linkcolor=blue,
    citecolor=blue,
    urlcolor=blue,
    filecolor=blue
}

\begin{document}

\title{3D Convolutional Neural Networks Image Registration Based on Efficient Supervised Learning from Artificial Deformations}

\author{Hessam~Sokooti,
		Bob~de~Vos,
		Floris~Berendsen,
		Mohsen~Ghafoorian,
		Sahar Yousefi, \\
        Boudewijn~P.F.~Lelieveldt,
        Ivana~I\v sgum,
        and~Marius~Staring~
\thanks{Hessam~Sokooti, 
		Floris~Berendsen,
		Sahar Yousefi,
        Boudewijn P.F.~Lelieveldt
        and \mbox{Marius~Staring} are with the Division of Image Processing, Department of Radiology, Leiden University Medical Center, Leiden, The Netherlands. Boudewijn P.F.~Lelieveldt and Marius~Staring are also with the Delft University of Technology, Delft, The Netherlands.
Bob~de~Vos and Ivana~I\v sgum are with the Image Sciences Institute, University Medical Center Utrecht, Utrecht, The Netherlands.
Mohsen~Ghafoorian is with TomTom, Amsterdam, The Netherlands}
\thanks{This work is financed by the Netherlands Organization for Scientific Research (NWO), project 13351. Dr. M.E.~Bakker and J.~Stolk are
acknowledged for providing a ground truth for the SPREAD study data
used in this paper. We would like to thank Dr. R.~Castillo and T.~Guerrero for providing the DIR-Lab databases.}
        }

\maketitle
%
%\author{Hessam Sokooti, Bob de Vos, Floris Berendsen, Mohsen Ghafoorian, \\ 
%Boudewijn P.F. Lelieveldt, Ivana I\v sgum, and Marius Staring\\
%Leiden University Medical Center, Leiden, the Netherlands\\
%University Medical Center Utrecht, Utrecht, the Netherlands\\
%TomTom, Amsterdam, the Netherlands\\
%Delft University of Technology, Delft, the Netherlands 
%}

% Media Template
%\author[LUMCadd]{Hessam Sokooti}
%\author[UMCadd]{Bob de Vos}
%\author[LUMCadd]{Floris Berendsen}
%\author[LUMCadd]{Sahar Yousefi}
%\author[TomTomadd]{Mohsen Ghafoorian}
%\author[LUMCadd,Delftadd]{Boudewijn P.F. Lelieveldt}
%\author[UMCadd]{Ivana I\v sgum}
%\author[LUMCadd,Delftadd]{Marius Staring}
%\address[LUMCadd]{Leiden University Medical Center, Leiden, The Netherlands}
%\address[UMCadd]{University Medical Center Utrecht, Utrecht, The Netherlands}
%\address[TomTomadd]{TomTom NV, Amsterdam, The Netherlands}
%\address[Delftadd]{Delft University of Technology, Delft, The Netherlands}

\begin{abstract}

We propose a supervised nonrigid image registration method, trained using artificial displacement vector fields (DVF), for which we propose and compare three network architectures. The artificial DVFs allow training in a fully supervised and voxel-wise dense manner, but without the cost usually associated with the creation of densely labeled data. 
We propose a scheme to artificially generate DVFs, and for chest CT registration augment these with simulated respiratory motion.
The proposed architectures are embedded in a multi-stage approach, to increase the capture range of the proposed networks in order to more accurately predict larger displacements. The proposed method, RegNet, is evaluated on multiple databases of chest CT scans and achieved a target registration error of \mbox{$2.32 \pm 5.33$ mm} and \mbox{$1.86 \pm 2.12$ mm} on SPREAD and DIR-Lab-4DCT studies, respectively. The average inference time of RegNet with two stages is about 2.2 s.

\end{abstract}

%\end{frontmatter}

\section{Introduction}
%\pdfmarkupcomment[markup=Highlight,color=CommentColor]{database}{is it better to remove this citation and write the url}

Image registration is the process of aligning images and has many applications in medical image analysis. Generally, image registration casts to an optimization problem of minimizing a predefined handcrafted intensity-based dissimilarity metric over a transformation model. Both the dissimilarity metric and the transformation model need to be selected and tuned in order to achieve high quality registration performance. This task is time-consuming and there is no guarantee that the selected dissimilarity model fits with new images. 

%\pdfmarkupcomment[markup=Highlight,color=CommentColor]{Suggested Structure}{1. Conventional learning papers 
%2. very early and general papers. 
%3. Unsupervised 
%4. Supervised 
%5. Reinforcement learning
%6. Model-based conventional }

General learning-based techniques have been used in several registration papers. \citet{guetter2005learning} incorporated a prior learned joint intensity distribution to perform a non-rigid registration. \citet{jiang2008learning} selected and fused a large number of features instead of using only one similarity metric. \citet{hu2017learning} leveraged regression forests to predict an initial DVF. In terms of predicting registration accuracy \citet{muenzing2012supervised} casted this task to a classification problem and extracted several local intensity-based features, which are fed to a two-stage classifier. \citet{sokooti2016accuracy, sokooti2019quantitative} extracted some intensity-based and registration-based features, then by using regression forests estimated the local registration error.

In recent years, CNNs have also been utilized in the context of image registration. \citet{miao2016cnn} used CNNs for rigid-body transformations. \citet{yang2016fast} trained a CNN to predict the initial momentum of a 3D LDDMM registration.  \citet{cao2017deformable} generated a multi-scale similarity map and utilized it to predict the DVF. \citet{simonovsky2016deep} proposed a CNN-based similarity metric for multi-modal registration. Their training samples were a set of aligned images as the positive cases and a set of manually deformed images as the negative cases.

In the unsupervised deep learning approaches, \citet{de2017end, de2019deep} for the first time used normalized cross correlation (NCC) of the fixed and moving image as a loss function. Later \citet{balakrishnan2018unsupervised,  ferrante2018adaptability} used the same loss to train their network. \citet{mahapatra2018joint} combined NCC with other similarity metrics such as the Dice overlap metric over the labeled images. \citet{elmahdy2019adversarial} eutilized an adversarial training based on the segmentation maps in addition to the NCC loss.
\citet{sheikhjafari2018unsupervised} and \citet{dalca2018unsupervised} employed the mean squared intensity difference, which was applied to mono-modal image registration. \citet{hu2018label} proposed a loss function that calculates cross entropy over the smoothed segmentation maps, which was applied to multi-modal images. A drawback to use conventional similarity metrics is that these similarity metrics are not perfect and might not fit in all images.

In the supervised approaches, for the first time \citet{sokooti2017nonrigid} generated artificial DVFs with different frequencies to train a CNN architecture.
\citet{rohe2017svf} proposed to build reference DVFs which were obtained by performing registration over segmented regions of interest. \citet{fan2018adversarial} proposed a ground truth based on the GAN network. The implicit ground truth is assigned using the negative cases derived from the generator network while the positive cases are synthetically made by perturbing the original images. \citet{eppenhof2018deformable} constructed a small set of images by applying a random DVF. In the model-based methods \citet{uzunova2017training} proposed statistical appearance models to be used for data augmentation. \citet{hu2018adversarial} utilized biomechanical simulations to regularize their network.

In several articles, reinforcement learning is used \citep{ma2017multimodal, krebs2017robust, onieva2018diffeomorphic}. An artificial agent is trained by making a statistical deformation model from training data. However, this approach is still iterative and might be slow at inference time.

Conventionally, in quality assessment of registration, manually selected landmarks or manually segmented regions are used. However, utilizing them as a gold standard in training has some drawbacks. With manually segmented regions, several measurements like Dice and mean surface distance can be calculated, but there is no direct correlation between Dice and the true DVF in all voxels of the image. The drawback of using landmarks as a gold standard \cite{sokooti2016accuracy, sokooti2019quantitative} is that the numbers of landmarks usually is not enough to estimate a continuous gold standard DVF for the whole image. 

In this paper, instead of using a transformation model, we directly predict the displacement vector field (DVF). The convolutional neural network (CNN) implicitly learns the dissimilarity metric. The current paper is a large extension of the work first presented in \citet{sokooti2017nonrigid}.
We present more ways to construct sufficiently realistic synthetic DVFs. 
The network design is greatly enhanced by increasing the capture range in order to more accurately predict larger DVFs. A multi-stage approach is also proposed to overcome this issue. The evaluation is performed on the SPREAD database as well as on the public DIR-Lab databases. The proposed method is capable to be trained on any database without needing any manual ground truth.

\section{Methods}

\subsection{System overview}

\begin{figure}[tb!]
\centering
\includegraphics[width=1\columnwidth, trim={10 80 10 85},clip]{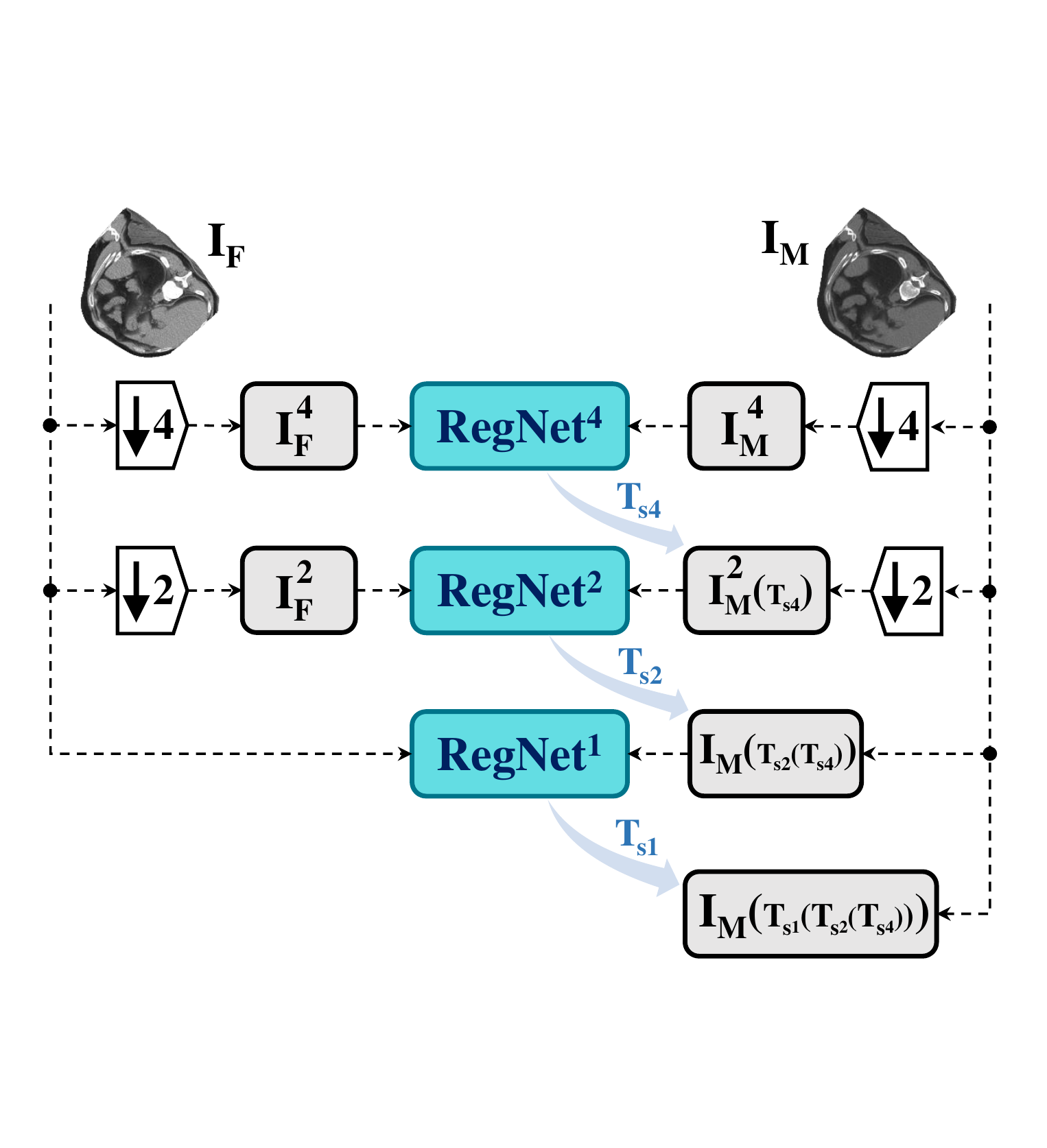}
\caption{Block diagram of the proposed system. The initial inputs of the system are fixed and moving images down-scaled by a factor of four ($\,\DownArrow[10pt][>=latex, black, thick]\:$4). Three RegNets process the input images over three stages (4, 2, 1) and generate the final output ${\bm{T}(\bm{x}) = \bm{T_{s1}}\big(\bm{T_{s2}}\big(\bm{T_{s4}(x)}\big)\big)}$. }\label{fig:BlockDiagram}
\end{figure}

A block diagram of the proposed system is given in Fig. \ref{fig:BlockDiagram}. The inputs of the system are a fixed image $I_{F}$ and a moving image $I_{M}$. Similar to the conventional registration methods, a multi-stage approach is employed. The registration blocks RegNet$^4$ and RegNet$^2$ perform on the down-scaled images with a factor of 4 and 2, respectively. The inputs of the final registration block RegNet$^1$ are original resolution images. The output of the system is a predicted DVF of transforming the moving image to the fixed image which is defined as ${\bm{T}(\bm{x}) = \bm{T_{s1}}\big(\bm{T_{s2}}\big(\bm{T_{s4}(x)}\big)\big)}$.

\subsection{Network architecture} \label{sec:methods:network}

We propose three network architectures for the RegNet design. 
The first two architectures are patch-based, and predict the DVF for a local neighborhood. These two networks are more complex and occupy a relatively large amount of GPU memory. The third architecture is based on a more simple U-Net design \citep{ronneberger2015u} with fewer network weights, and is capable of registering entire images (not patches), but down-scaled, within the memory limits of current GPUs. This last architecture is considered a candidate for the first resolution (RegNet\textsuperscript{4}), while the others are considered for the second and third resolution (RegNet\textsuperscript{2} and RegNet\textsuperscript{1}). In Section \ref{sec:experiments} we compare these architectures and combinations thereof.

%
%
%is called ``smaller output'' in which contextual information are used to predict the DVF for a center region of the inputs. These designs (U-net-filled and decimation) are more complex and they can be utilized mostly in later stages of registration  i.e RegNet\textsuperscript{2,1}. 

%One of the reason of designing of these patch-based network is the limitation of GPU memory (typically 12 GB max). For the second type ``Equally input-output size'' we propose a more simple design based on U-Net \citep{ronneberger2015u}, which allows us to give the entire (down-scaled) image to the network. This design can be used especially in RegNet\textsuperscript{4}. Later, we compare the combination of the proposed network in Section \ref{sec:experiments}.

The networks have some settings in common. All convolutional layers use batch normalization
% \citep{ioffe2015batch} 
and ReLu activation
%\citep{nair2010rectified}, 
,except for the last two layers of the U-Net design and the last three layers
of the patch-based designs, where ELu activation is used to improve the regression accuracy. The last layer of all architectures does not use batch normalization nor an activation function. 
The Glorot uniform initializer
% \citep{glorot2010understanding}
is used for all convolutional layers except for the trilinear upsampling, in which a fixed trilinear kernel is utilized.
%\citep{shelhamer2016fully}. 
The three architectural designs are given in Fig. 2. The details are:

\subsubsection{U-Net (U)} U-Net is one of the most common designs used in medial image segmentation
%\citep{litjens2017survey}. 
The proposed modified design has an input size and output size of \mbox{$125\times125\times125$} voxels. This architecture is only used for the sub-block \mbox{RegNet\textsuperscript{4}}, i.e. CNN-based registration is applied to down-scaled images with a factor of four. The proposed design is given in Fig. \ref{fig:Network:unet}. This relative simple design has 232,749 trainable parameters.

\subsubsection{Multi-View (MV)} 
In this design, different scales are created by using conventional decimation by convolving the inputs with fixed B-spline kernels, which is similar to \cite{sokooti2017nonrigid} and \cite{kamnitsas2017efficient}. This design is relatively more memory efficient because of this multi-view approach. The proposed CNN architecture is visualized in Fig. \ref{fig:Network:decimaiton}. The input of the network is a pair of 3D patches of size \mbox{$105\times105\times105$} for the fixed and moving image. The network is then split into 3 pipelines: down-scaled with a factor of 4, a factor of 2, and the original resolution.
In order to save memory, the original resolution and the down-scaled version with a factor 2 are cropped to \mbox{$37\times37\times37$} and \mbox{$67\times67\times67$}, respectively. Decimation is done with the help of convolutions with a fixed B-spline kernel.
% \citep{parker1983comparison}. 
In the down-scaled factor 2 pipeline, a stretched B-spline kernel with size $7\times7\times7$ is used.
For down-scaling with a factor of 4, the B-spline kernel is stretched by a factor of 4, and has a size of $15\times15\times15$. 
Each pipeline continues with several convolutional layers with dilation of 1 or higher.
% \citep{yu2015multi}. 
The upsampling layers ensure spatial correspondence of all three pipelines. Finally, all pipelines are merged together followed by three more convolutional layers. The network gives three 3D outputs of size $21\times21\times21$ corresponding to the displacement in $x$, $y$ and $z$ direction. The total number of parameters in this design is 1,201,353.

\subsubsection{U-Net-Advanced (Uadv)}
\begin{figure}[tb]
\begin{subfigure}{1\columnwidth}
	\centering
	\includegraphics[width=1\columnwidth, trim={10 125 260 0},clip]			{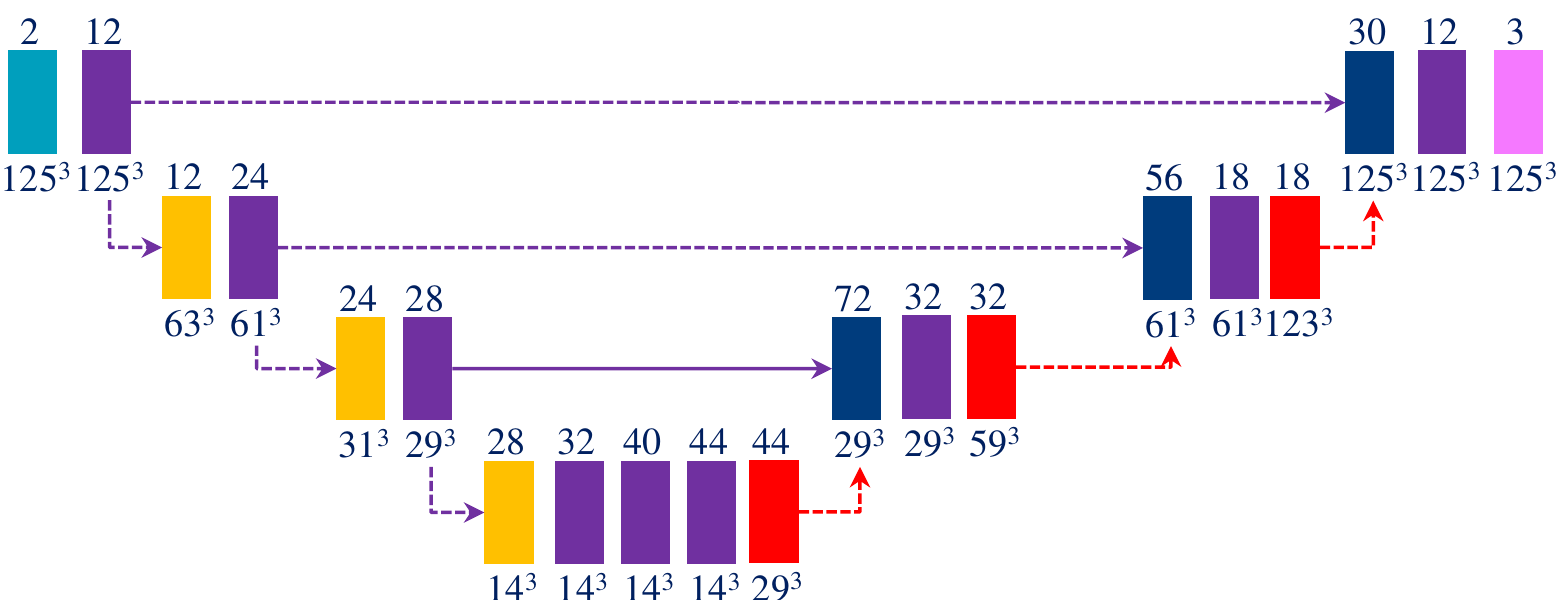}
	\caption[]{U-Net (U)}
	 \label{fig:Network:unet}
\end{subfigure}
\begin{subfigure}{1\columnwidth}
\centering
	\includegraphics[width=1\columnwidth, trim={10 105 260 0},clip]{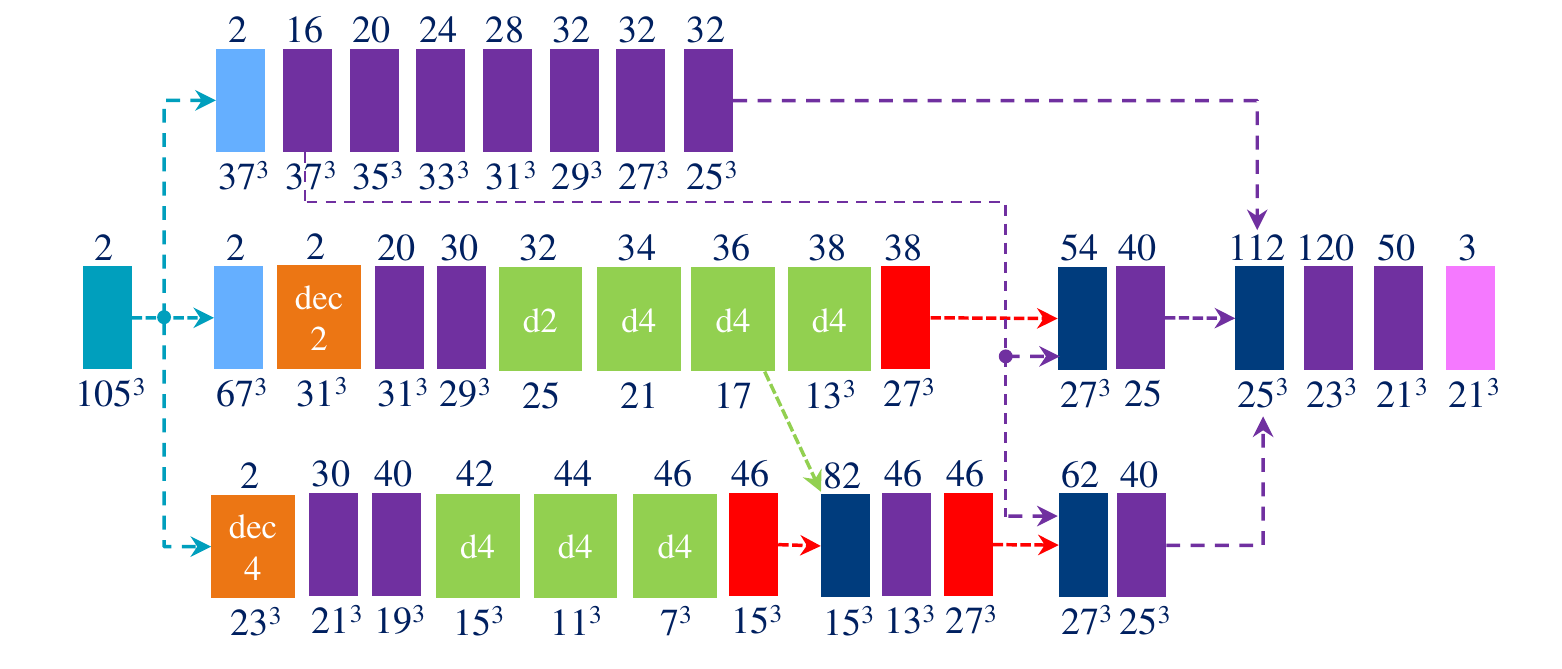} 
	\caption[]{Multi-view (MV)}
	\label{fig:Network:decimaiton}
\end{subfigure}
\begin{subfigure}{1\columnwidth}
\centering
\includegraphics[width=1\columnwidth, trim={10 40 260 0},clip]{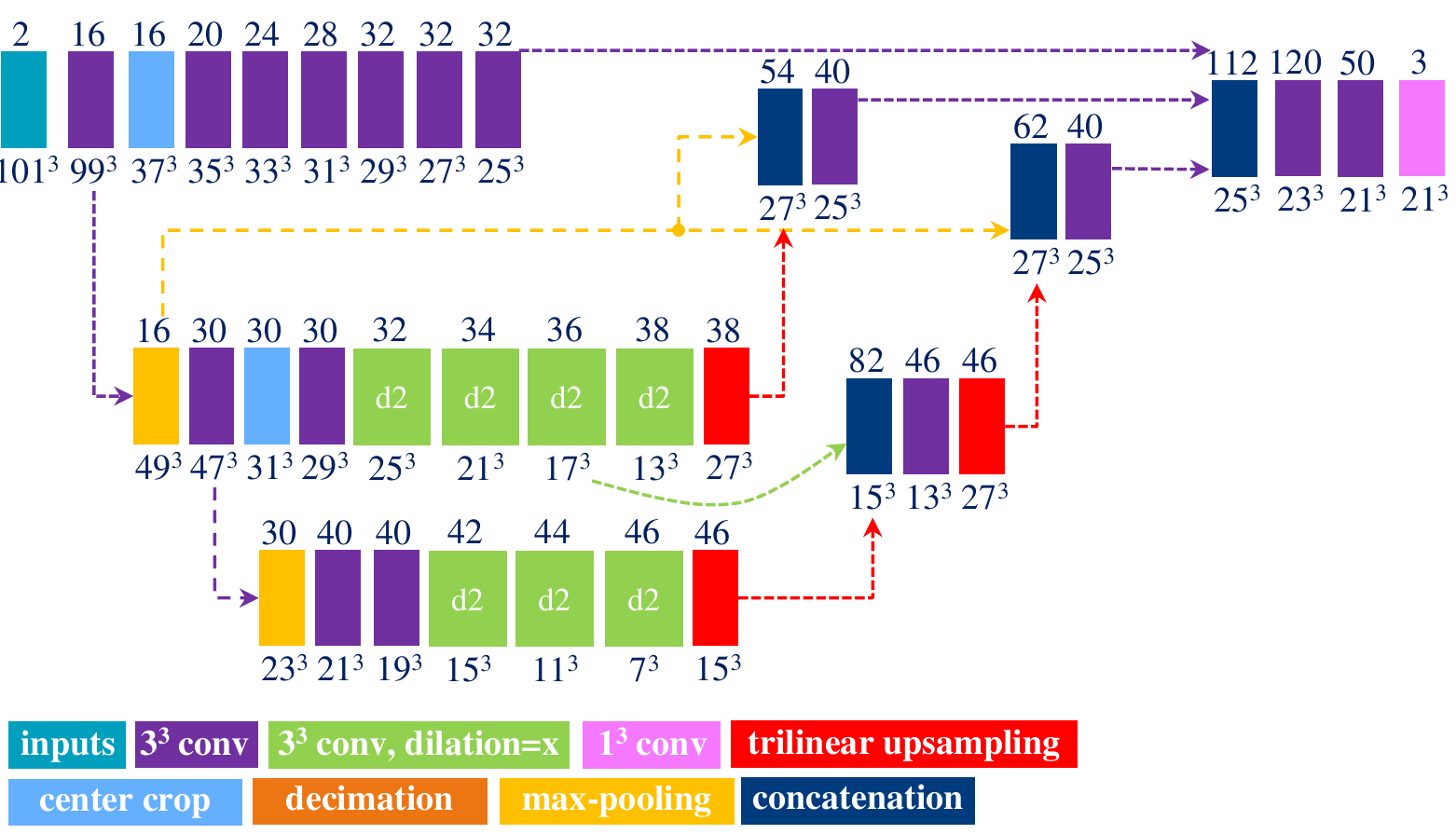}
\caption[]{U-Net-advanced (Uadv)}
\label{fig:Network:maxpool}
\end{subfigure}
\caption{RegNet designs: The inputs of the U-Net design are entire down-scaled images. However, in the Multi-view and U-Net-advanced architectures the output size is smaller than the input size and can be trained in a patch-based manner.}\label{fig:Network}
\end{figure}

This proposed architecture is again a patch-based one but using a max-pooling technique instead of a decimation method. The global design is similar to the U-net architecture, but instead of simple shortcut connections, several convolutions are used for these connections. The proposed design is illustrated in Fig. \ref{fig:Network:maxpool}. The network starts with a convolutional layer to extract several low-level features from the images before any max-pooling. The size of the inputs and output are \mbox{$101\times101\times101$} and \mbox{$21\times21\times21$}. The total number of parameters in this design is 1,420,701.

\subsection{Artificial generation of DVFs and images}

In order to train a CNN, a considerable number of ground truth DVFs are needed. 
We take a moving image $I_M$ from the training set. The fixed image $I_F$ is created artificially by generating a DVF, applying the DVF to the moving image resulting in $I_{F}^{\mathrm{clean}}$, and adding artificial intensity models to finally obtain $I_F$.

\subsubsection{Artificial DVF} \label{sec:artificial_dvf}

We propose to generate three categories of DVFs, to represent the range of displacements that can be seen in real images:

\textbf{single frequency:}
The first category consists of DVFs having one or more local displacements of only one spatial frequency. They are generated as follows: Create an empty B-spline grid of control points with a spacing of \mbox{$s$ mm}; Assign random values to the grid of control points and smooth it with a Gaussian kernel; Resample the B-spline grid to obtain the DVF; Normalize the DVF linearly to be in the range $[-\theta, +\theta]$ along each axis.

\textbf{mixed frequency:}
In this category, two different spatial frequencies are mixed together as follows:
Create a single frequency DVF similar to the previous category; Create a random binary mask and multiply it with the single frequency DVF. Finally, smooth the DVF with a Gaussian kernel with a standard deviation of $\sigma_B$. $\sigma_B$ is chosen relatively small to generate a higher spatial frequency in comparison with the smooth filled region;
By varying the $\sigma_B$ value and $s$ in the filled DVF, different spatial frequencies will be mixed together.

%Extract edges with the Canny edge detection method with a low and high threshold value of 50 and 100, respectively;
%Create an initial DVF by this binary image (simply repeating it three times);
%Set several voxels randomly to zero;
%Dilate the binary image for $N_d$ iterations using a structuring element varying between 6-connected and 27-connected;
%Fill the above binary image with random smooth values (a DVF generated from the single frequency method);

\textbf{respiratory motion:} We simulate respiratory motion with three components similar to \cite{hub2009stochastic} as follows:
Expansion of the chest in the transversal plane with a maximum scaling factor of 1.12;
Transition of the diaphragm in cranio-caudal direction with a maximum deformation of $\theta$;
Random deformation using the single frequency method. In order to locate the diaphragm, an automatically detected lung mask is used.

\textbf{identity:} 
This category comprises only identity DVFs. Later, when creating the artificially deformed image, intensity augmentations will be added to the deformed image. Thus, the network will be capable of detecting no motion, while the intensity values might have changed slightly.

\begin{figure}[tb]
\centering
\includegraphics[width=1\columnwidth, trim={15 25 15 20},clip]{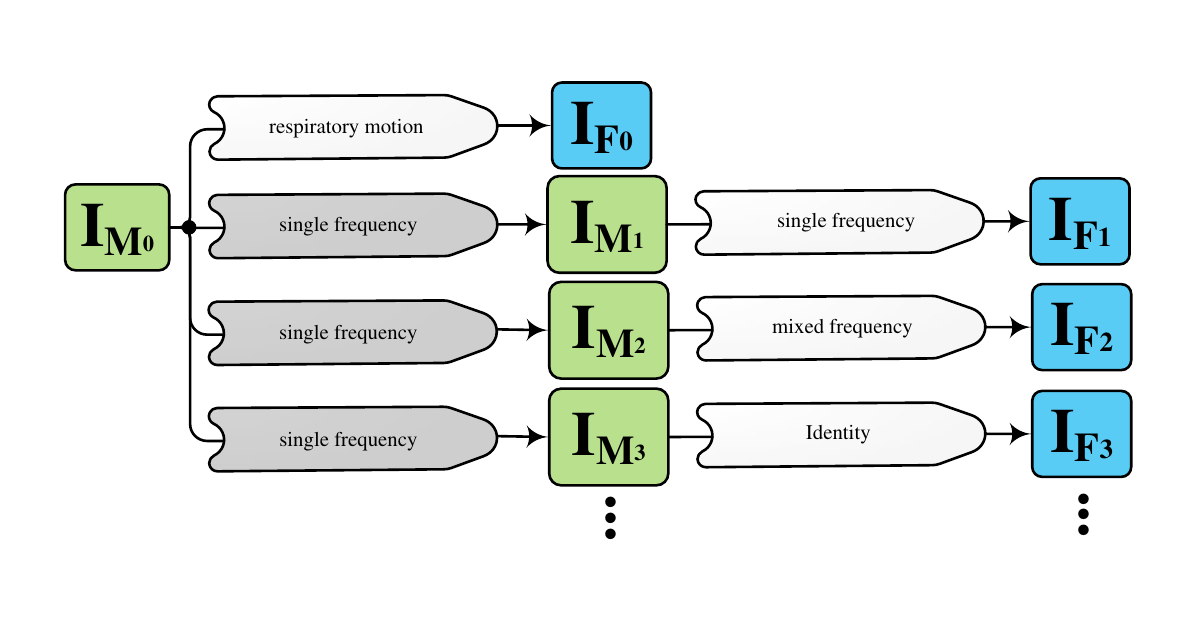}
\caption{The generation of training pairs from a single input image $I_{M0}$. The input image is deformed slightly using the single frequency category, with the ``lowest'' settings (see Table \ref{tb:DVF_freq}), to generate moving images $I_{Mi}$. These are then each deformed and post-processed multiple times using all categories to generate fixed images $I_{Fi}$.
}\label{fig:training}
\end{figure}

\subsubsection{Artificial intensity models} \label{sec:method:ag:artificial_deformed_image}
We propose two intensity models to be applied on the fixed images:

\textbf{Sponge intensity model:} By assuming mass preservation over the lung deformation, a dry sponge model \citep{staring2014towards} is added to deformed image:

\begin{align}
\bm{I_F}(x) = \bm{I_{F}^{\mathrm{clean}}}(x)[\bm{J_T}(x)^{-1}],
\label{eq:sponge}
\end{align}
where $\bm{J}$ denotes the determinant of the Jacobian of the transformation.

\textbf{Gaussian noise:} A Gaussian noise with a standard deviation of $\sigma_N=5$ is added to the deformed image in order to achieve more accurate simulation of real images. 

\subsubsection{Extensive pair generation} \label{sec:extensive_generation}
for each single image in the training set, potentially a large number of artificial DVFs can be randomly generated. However, if this image is to be re-used for multiple DVFs, then for many training pairs we have the moving image unaltered. To tackle this problem, we also generate deformed versions of the original image (gray single frequency blocks in Fig. \ref{fig:training}). A schematic design of utilizing artificial image pairs is depicted in \mbox{Fig. \ref{fig:training}}. In this approach, the original image is only used once to generate the artificial image $\mathrm{I_{F0}}$. Deformed versions of the original image  $I_{Mi}$ are used afterwards. Training pairs are thus $(I_{M0}, I_{F0}),(I_{M1}, I_{F1}), (I_{M2}, I_{F2}), ...$. The gray single frequency blocks in Fig. \ref{fig:training} have the same setting as single frequency ``lowest'' except that  $\sigma_N$ is set to 3 instead of 5. That is to avoid the accumulation of noise in the artificial images.

\begin{figure*}[tb]
\centering
\begin{subfigure}{0.32\textwidth}
	\includegraphics[width=0.625\linewidth, trim={0 -35 0 0}]{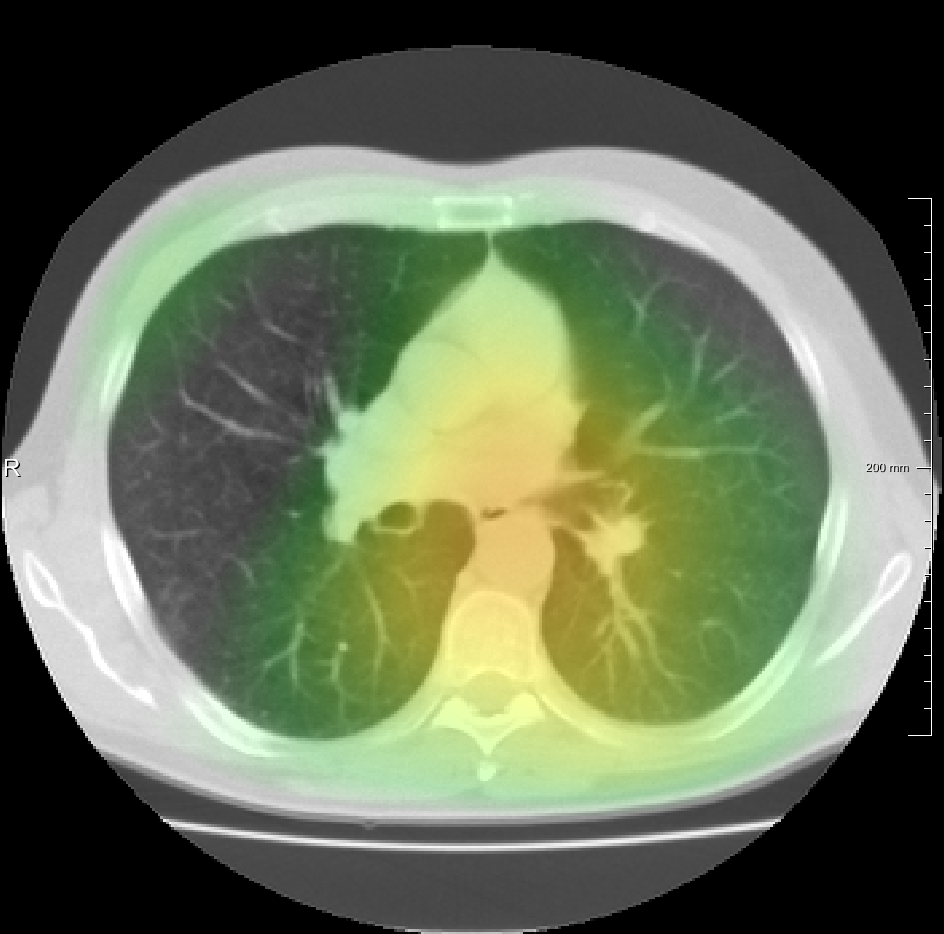}%	
	\includegraphics[width=0.375\linewidth, trim={45 60 10 50}, clip]{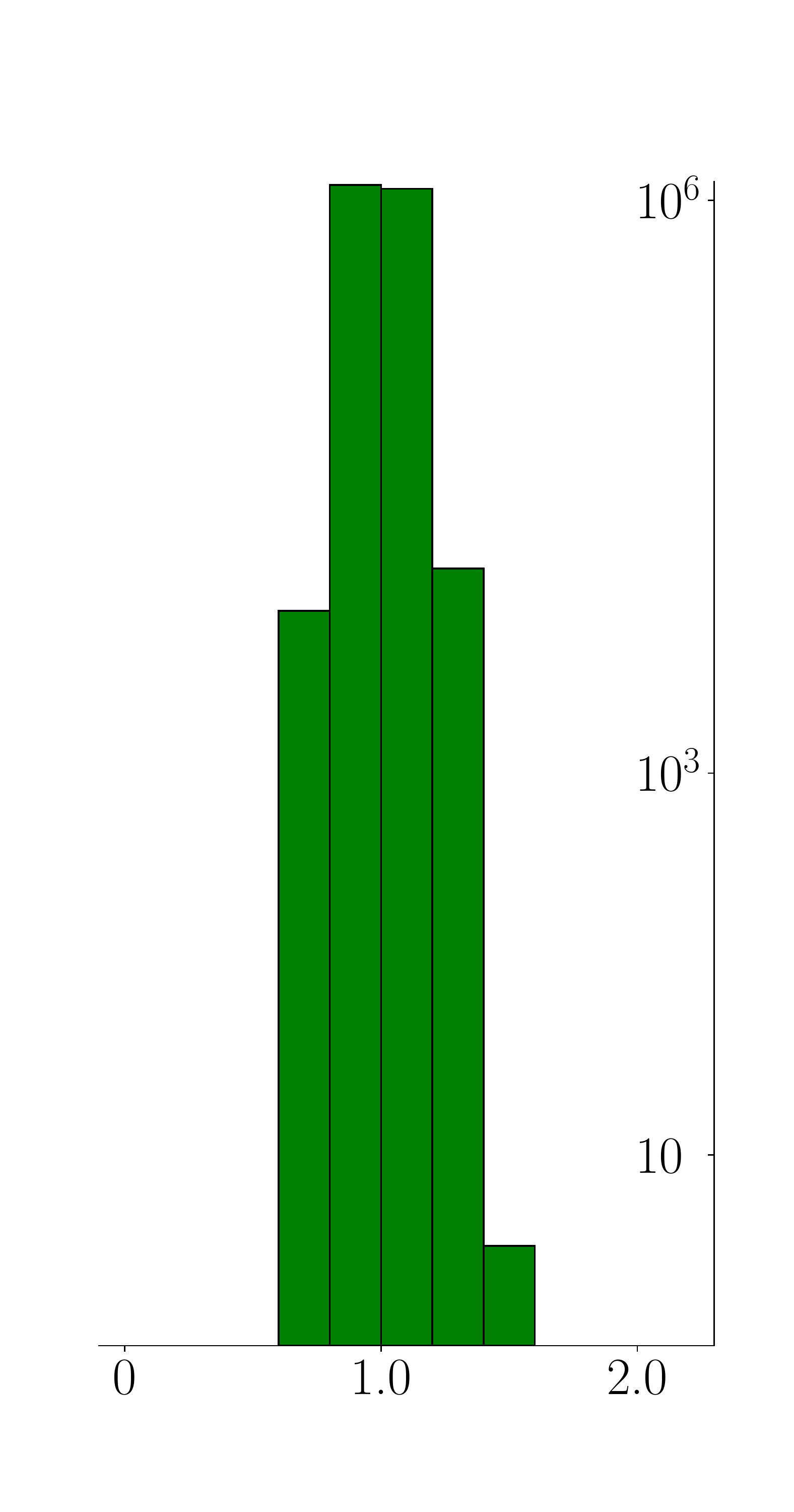}
		\caption[]{single frequency ``lowest''}
\end{subfigure}
\begin{subfigure}{0.32\textwidth}
	\includegraphics[width=0.625\linewidth, trim={0 -35 0 0}]{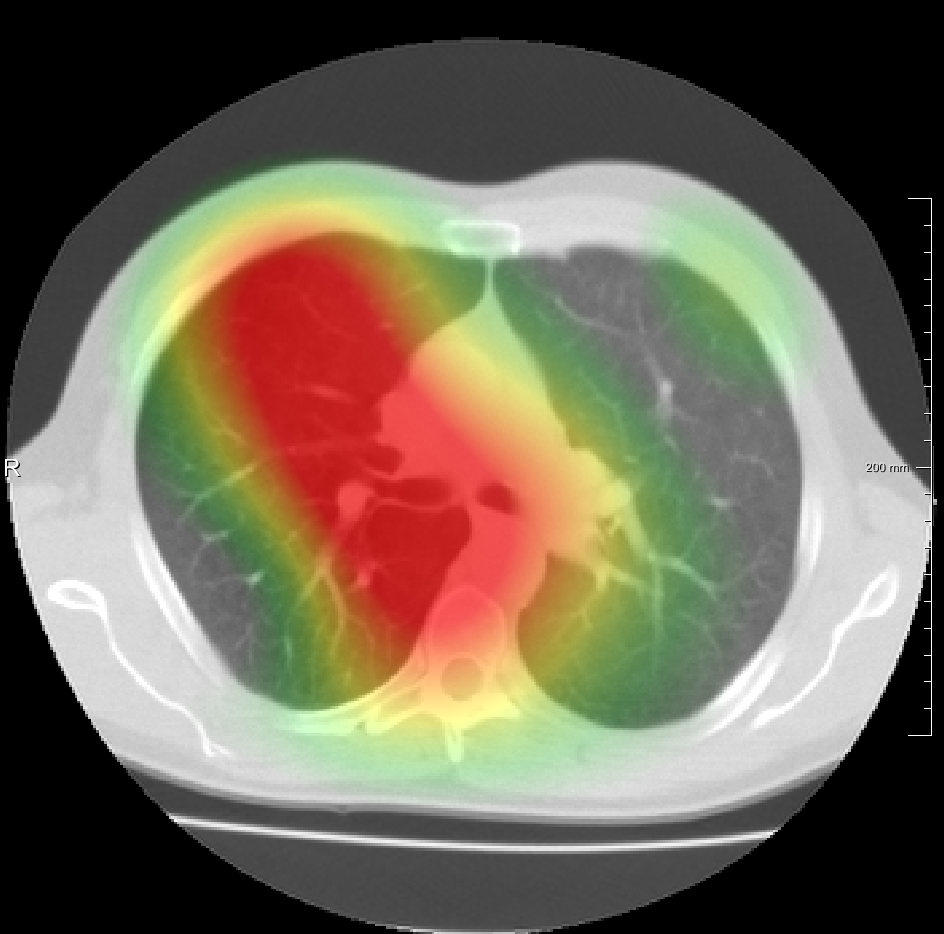}%	
	\includegraphics[width=0.375\linewidth, trim={45 60 10 50}, clip]{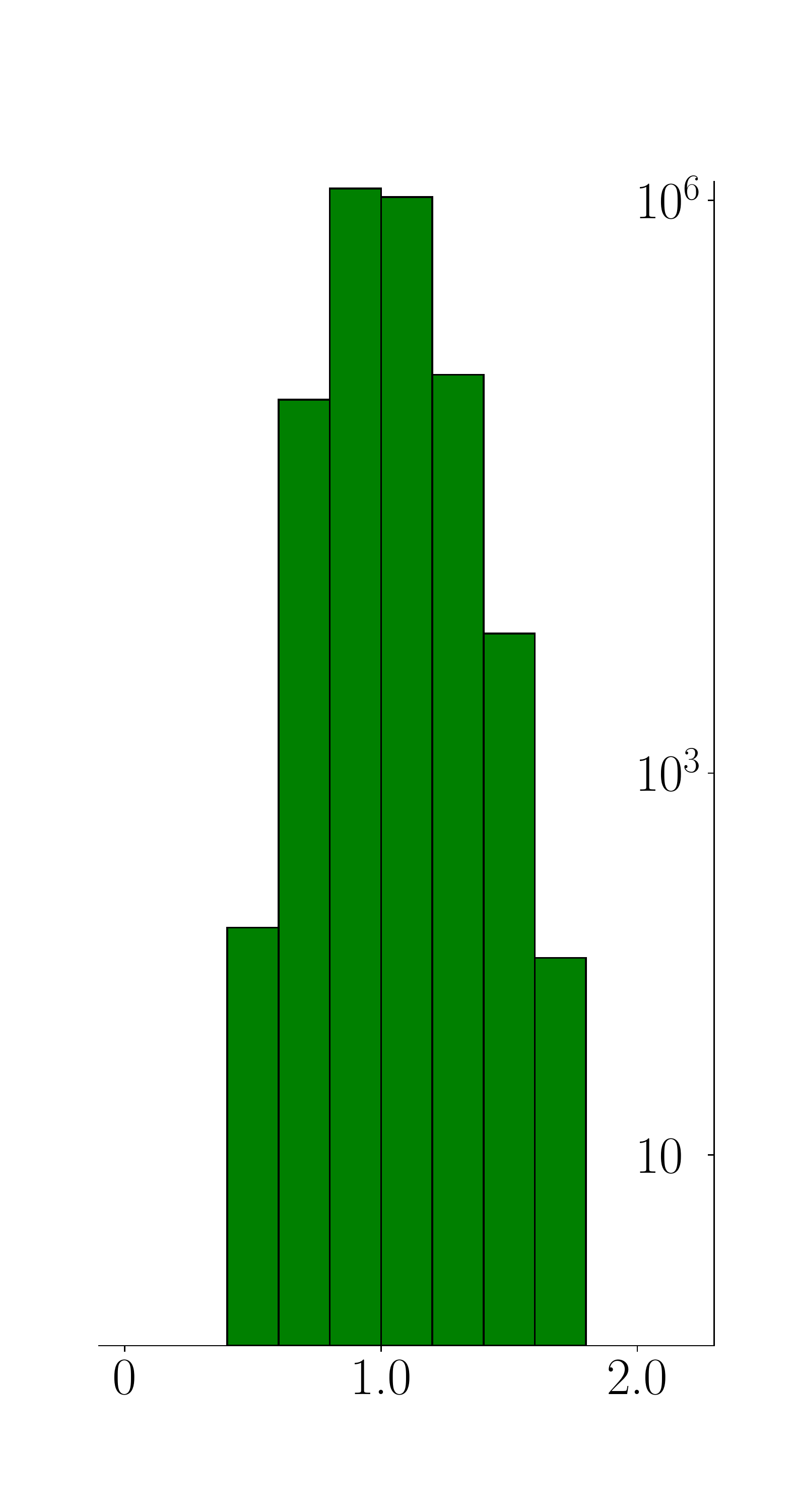}
		\caption[]{single frequency ``intermediate''}
\end{subfigure}
\begin{subfigure}{0.32\textwidth}
	\includegraphics[width=0.625\linewidth, trim={0 -35 0 0}]{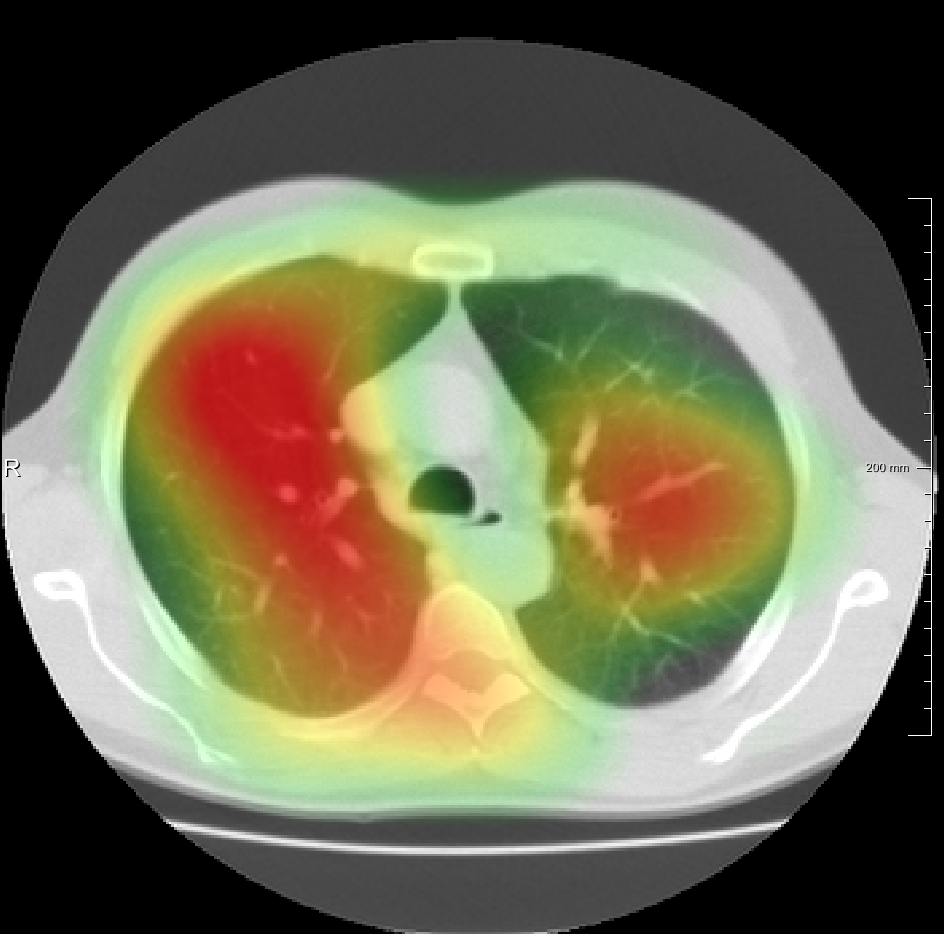}%	
	\includegraphics[width=0.375\linewidth, trim={45 60 10 50}, clip]{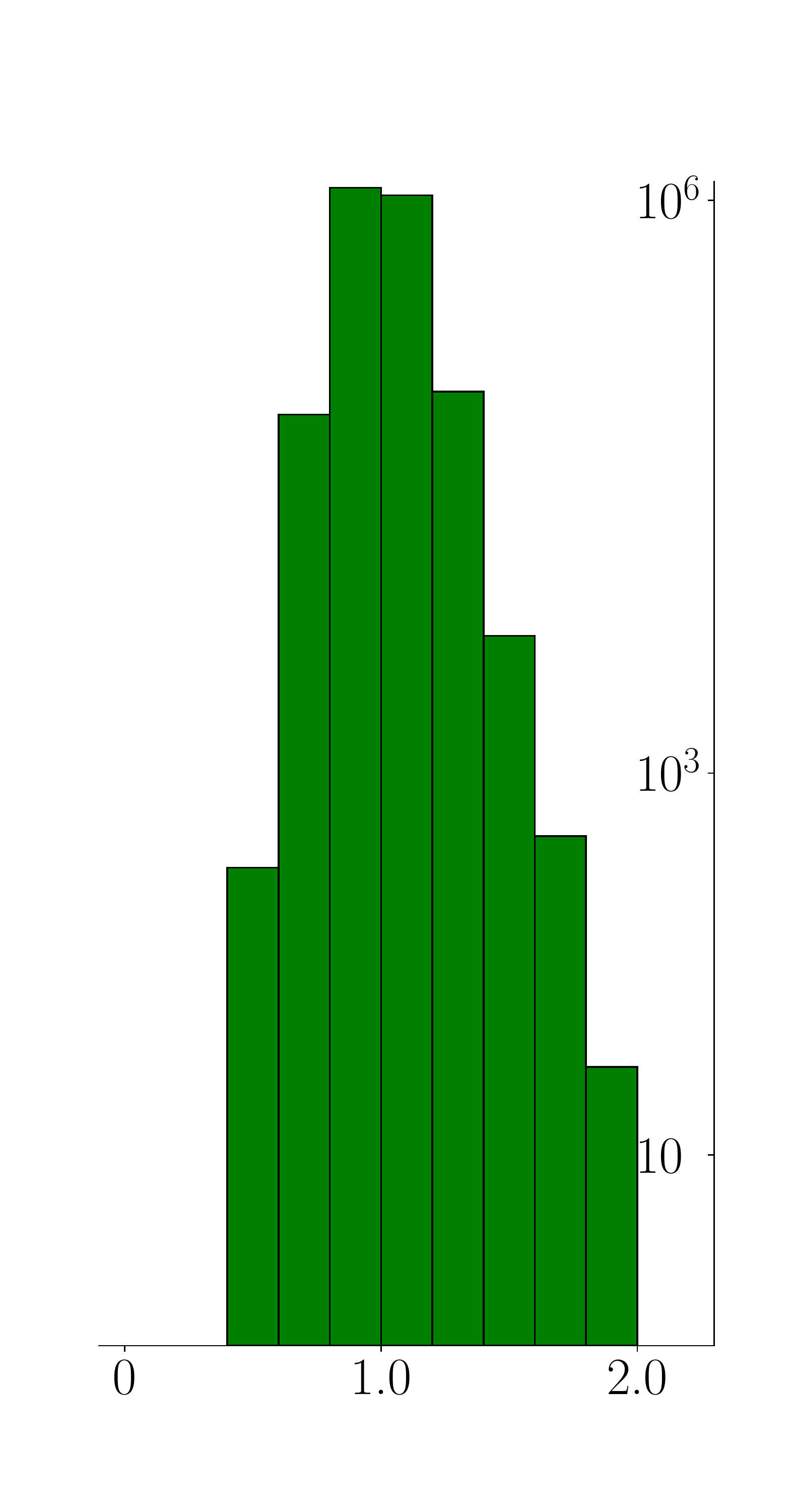}
		\caption[]{single frequency ``highest''}
\end{subfigure}
\begin{subfigure}{0.32\textwidth}
	\includegraphics[width=0.625\linewidth, trim={0 -30 0 0}]{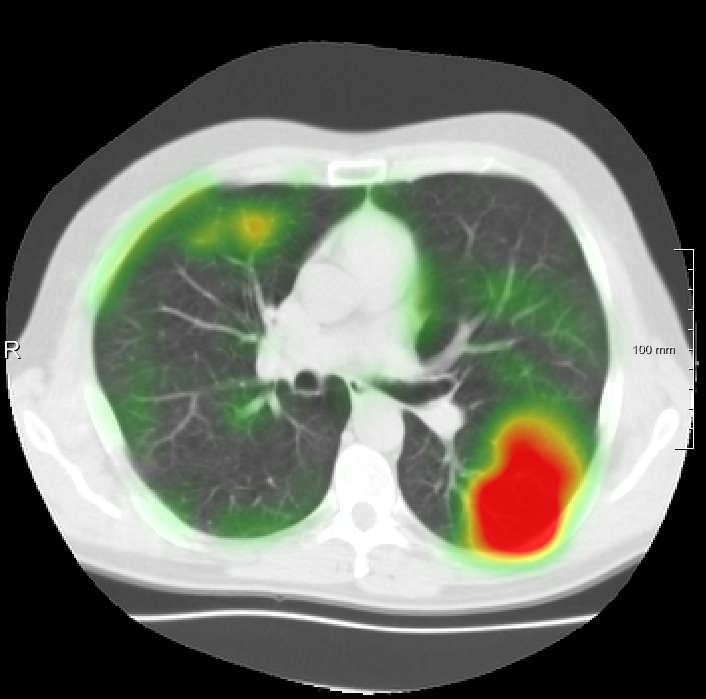}%	
	\includegraphics[width=0.375\linewidth, trim={45 60 10 50}, clip]{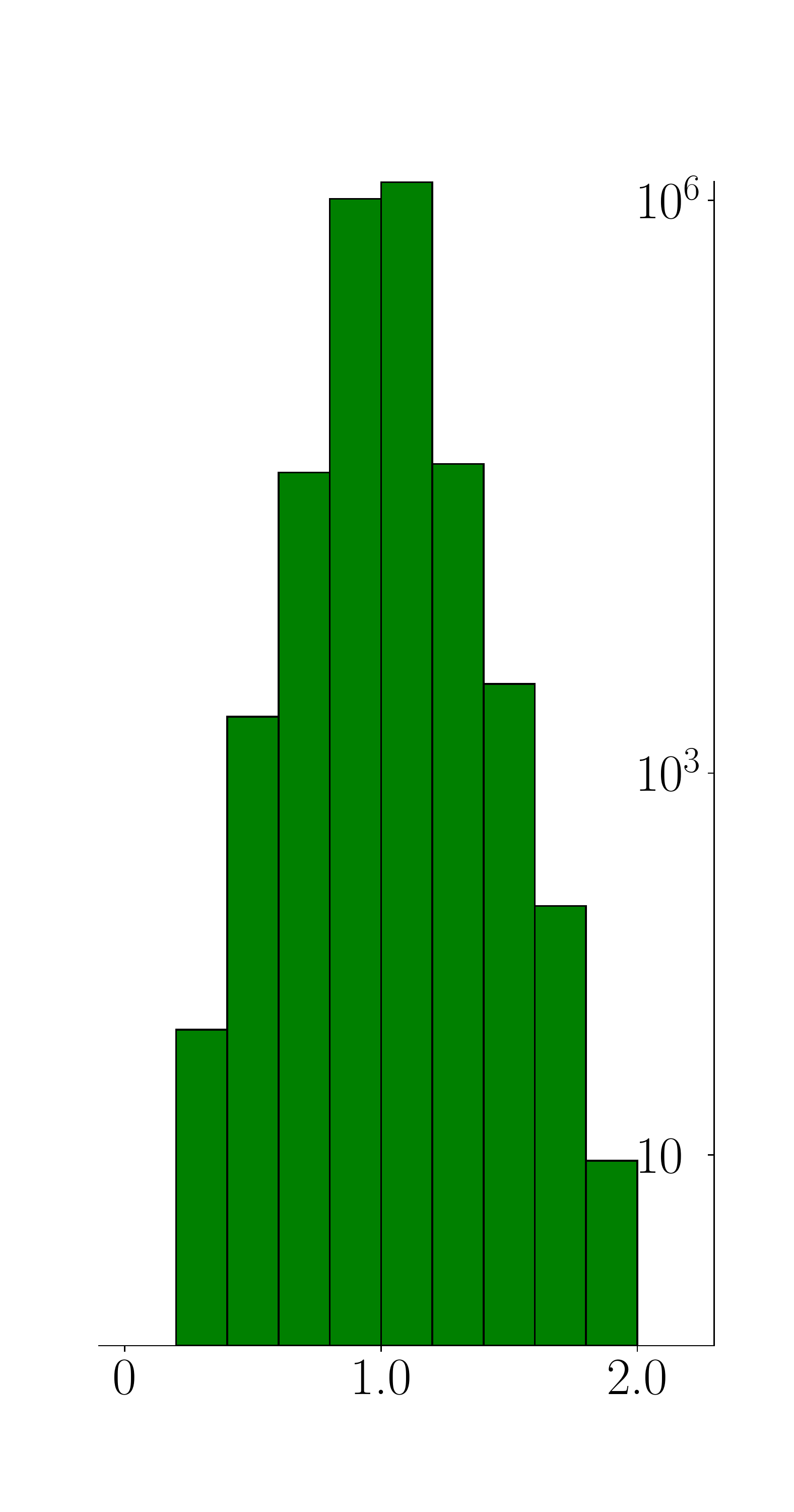}
		\caption[]{mixed frequency ``lowest''}
\end{subfigure}
\begin{subfigure}{0.32\textwidth}
	\includegraphics[width=0.625\linewidth, trim={0 -30 0 0}]{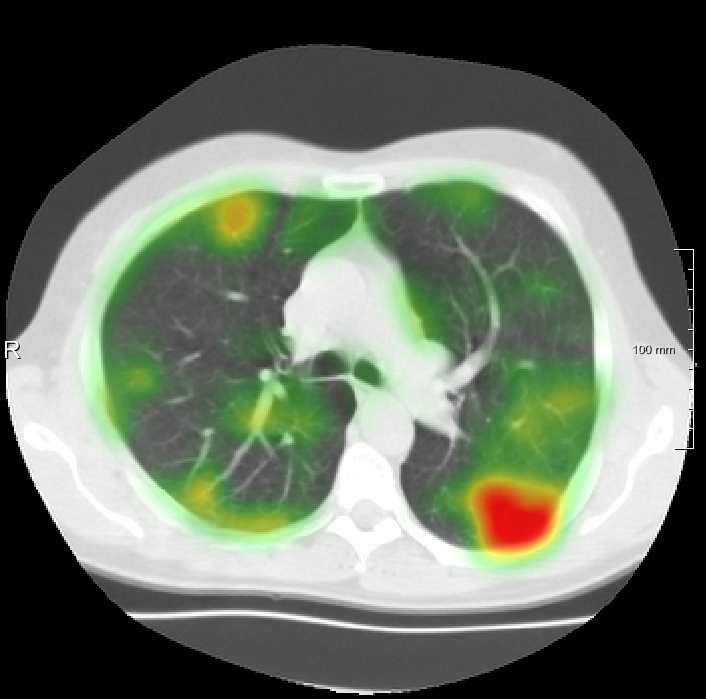}%	
	\includegraphics[width=0.375\linewidth, trim={45 60 10 50}, clip]{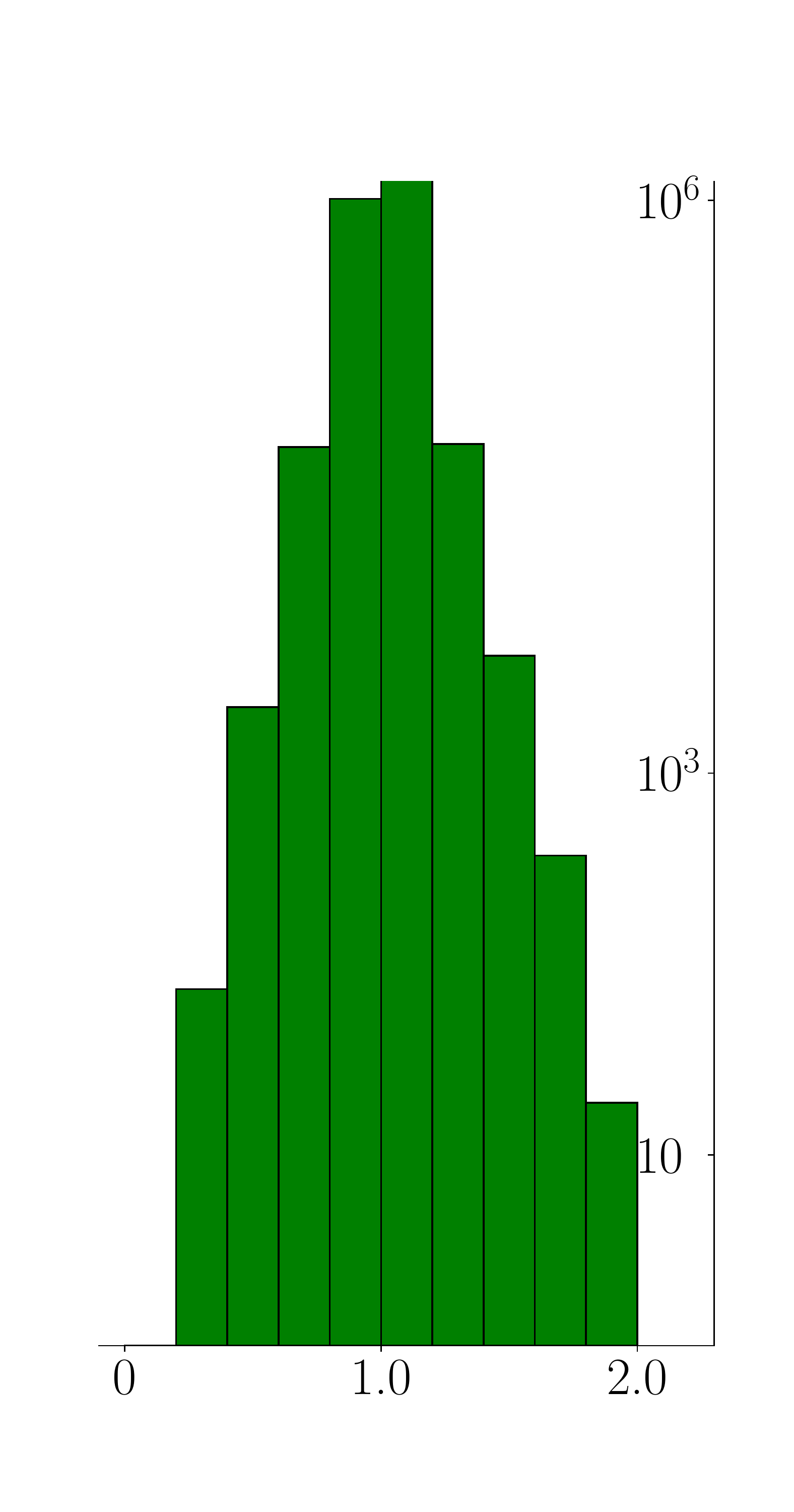}
		\caption[]{mixed frequency ``intermediate''}
\end{subfigure}
\begin{subfigure}{0.32\textwidth}
	\includegraphics[width=0.625\linewidth, trim={0 -30 0 0}]{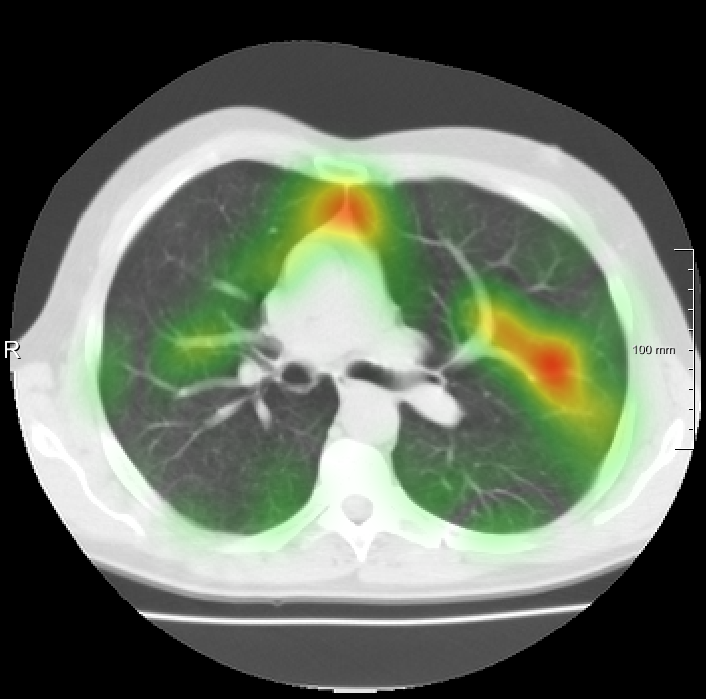}%	
	\includegraphics[width=0.375\linewidth, trim={45 60 10 50}, clip]{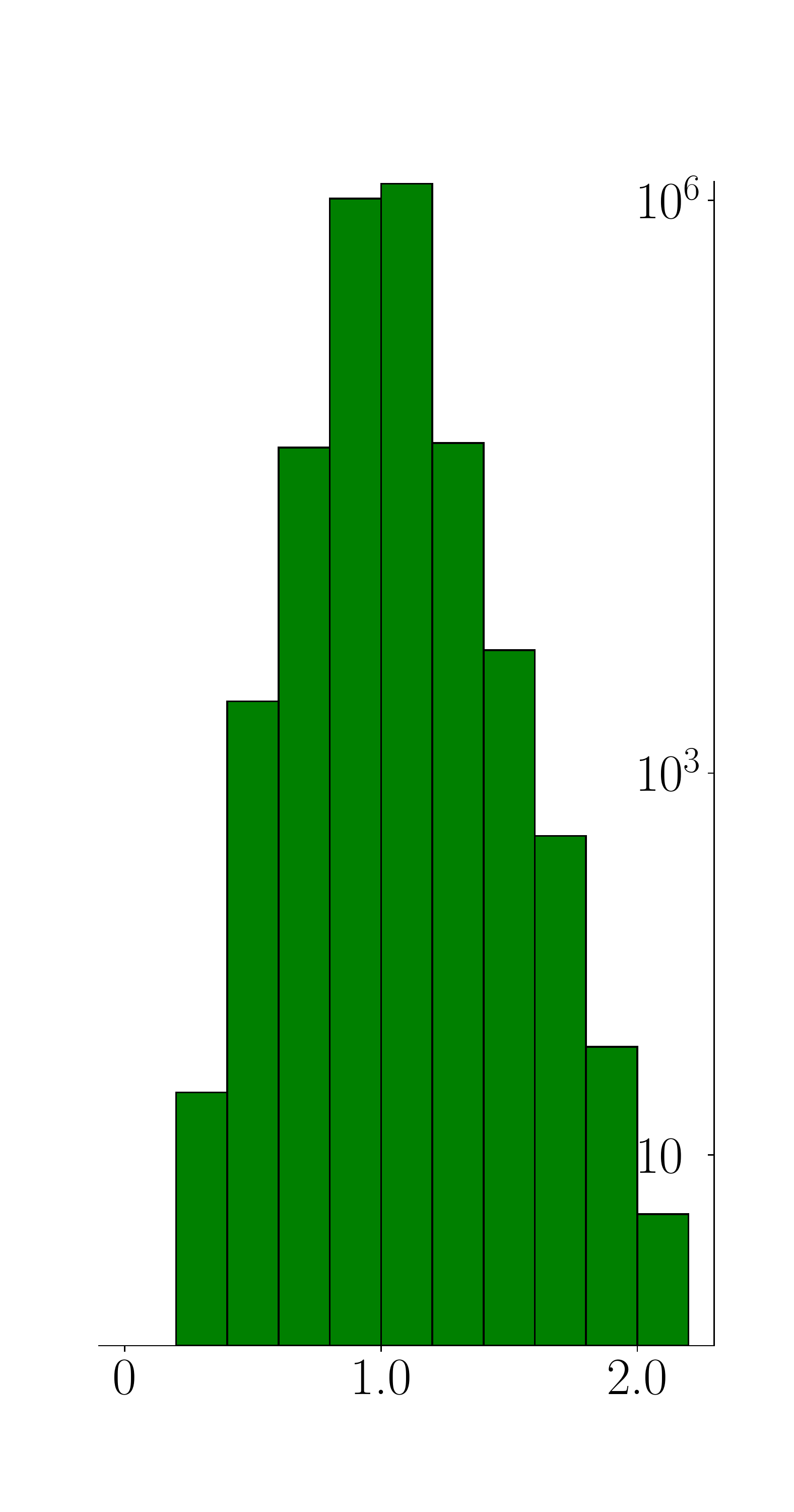}
		\caption[]{mixed frequency ``high''}
\end{subfigure}

%	\subfigure[mixed frequency ``high'']{\includegraphics[width=0.20\textwidth,trim={0 -30 0 0}]{mixed_type3.png}\includegraphics[width=0.12\textwidth ,trim={45 60 10 50}, clip]{Jac_Mixed_High2.pdf}}
\caption{Examples of heat maps of generated artificial DVFs overlayed on the deformed images. We show three of the five spatial frequencies defined in Table \ref{tb:DVF_freq}. The histogram of the Jacobian determinant of each DVF is shown next to the sample image. As the spatial frequency increases, the histogram is more spread.}\label{fig:synthetic_dvf}
\end{figure*}

In total we generate 14 basis types of artificial DVFs: 5 single frequency, 4 mixed frequency, 4 respiratory motion and 1 identity. The precise settings of the parameters are available in Table \ref{tb:DVF_freq} and examples are given in Fig. \ref{fig:synthetic_dvf}. The histograms of the Jacobians are also available in this figure. When the spatial frequency is increased, the Jacobian histograms will spread more, which shows that local relative volume changes are increased.
The value of $\theta$, the maximum artificial displacement along each axis, is chosen as 20, 15 and 7 for RegNet\textsuperscript{4}, RegNet\textsuperscript{2} and RegNet\textsuperscript{1}, respectively.

\begin{table*}[tb!]
	\centering
	\caption{DVFs with different spatial frequencies are obtained by varying the B-spline grid spacing $s$ and the standard deviation of the Gaussian kernel $\sigma_B$. The maximum deformation along each axis $\theta$ only varies for each stage. When the spatial frequency is increased, the Jacobian histograms will spread more, which shows that local relative volume changes are increased (Fig. \ref{fig:synthetic_dvf}). S, M and R indicates single frequency, mixed frequency and respiratory motion. }
%	\resizebox{\textwidth}{!}{
	\begin{tabular}{llccccc}
	\hline
	Parameter &artificial DVF &lowest  &low &intermediate &high &highest\\
	\hline

\multirow{3}{*}{$\theta$ (mm)}   &stage 1 &3 &7 &7 &7 &7  \\
		       &stage 2 &5 &15 &15 &15 &15  \\
   		       &stage 4 &7 &20 &20 &20 &20  \\

\rule{0pt}{4ex} \multirow{9}{*}{$s$ (mm)} & S$^1$ &[50, 50, 50] &[45, 45, 45] &[35, 35, 35] &[25, 25, 25] &[20, 20, 20]\\
& S$^2$ &[60, 60, 60] &[50, 50, 50] &[45, 45, 45] &[40, 40, 40] &[35, 35, 35]\\
& S$^4$ &[80, 80, 80] &[70, 70, 70] &[60, 60, 60] &[50, 50, 50] &[45, 45, 45]\\

& M$^1$ &[50, 50, 50] &[40, 40, 40] &[25, 25, 35] &[20, 20, 30] &\\
& M$^2$ &[60, 60, 60] &[50, 50, 40] &[40, 40, 80] &[35, 35, 80] &\\
& M$^4$ &[80, 80, 80] &[60, 60, 60] &[50, 50, 50] &[45, 45, 60] &\\

& R$^1$ &[50, 50, 50] &[45, 45, 45] &[35, 35, 35] &[25, 25, 25] &\\ 
& R$^2$ &[60, 60, 60] &[50, 50, 50] &[45, 45, 45] &[40, 40, 40] & \\
& R$^4$ &[80, 80, 80] &[70, 70, 70] &[60, 60, 60] &[50, 50, 50] & \\

\rule{0pt}{4ex} \multirow{3}{*}{$\sigma_B$} &M$^{1}$ &(5-10) &(5-10) &(5-10) &(5-10)  &  \\
 	&M$^{2}$ &(7-12) &(7-12) &(7-12) &(7-12)  &  \\
	&M$^{4}$ &(10-15) &(10-15) &(10-15) &(10-15)  &  \\
%$N_d$  &mixed$^{1,2,4}$ &200 &150 &150 &150 & \\

	\hline	
	\end{tabular}
%	}	
	\label{tb:DVF_freq}	
\end{table*}

\subsection{Optimization}
Optimization is done using the Adam optimizer
%\citep{kingma2015adam} 
with a learning rate of $0.001$. The loss function consists of two parts. The first part is the Huber loss,
%\citep{huber1964robust}
which minimizes the difference between the ground truth $\bm{T}$ and the predicted DVF $\bm{T'}$ of the RegNet. The second part is a bending energy (BE) regularizer \citep{rueckert1999nonrigid}, which ensures smoothness of the displacement field:
\begin{align}
 \mathcal{C}= \mathrm{Huber}\big(\bm{T}(\bm{x}), \bm{T'}(\bm{x})\big) + \gamma \cdot \mathrm{BE}\big(\bm{T'}(\bm{x})\big),
\label{eq:loss}
\end{align}
where the $\mathrm{Huber}$ loss is defined as:
\begin{align}
    \mathrm{Huber}(\bm{T}, \bm{T'}) = 
    \begin{cases}
       (\bm{T}-\bm{T'})^2,&  |\bm{T}-\bm{T'}| \leq 1,\\
       |\bm{T}-\bm{T'}|,& |\bm{T}-\bm{T'}| > 1
    \end{cases} 
\end{align}

%
%More details about the training of the networks are given in Section \ref{sec:NetworkDetails}.

\section{Experiments and results} \label{sec:experiments}
\subsection{Materials and ground truth} \label{title:Materials} 
Three chest CT scan datasets are used in this study: The SPREAD \citep{stolk2007progression}, the DIR-Lab-4DCT \citep{castillo2009framework} and the DIR-Lab-COPDgene dataset \citep{castillo2013reference}. 

In the SPREAD database, 21 pairs of 3D chest CT images are available with a baseline and a follow-up image in each pair. The follow-up images are taken after 30 months. Both images are acquired in the inhale phase. Patients in this study are aged between 49 and 78 years old. The size of the images is approximately $446\times 315 \times 129$ with a mean voxel size of \mbox{$0.78\times 0.78 \times 2.50$ mm}. About 100 well-distributed corresponding landmarks were previously selected \citep{staring2014towards} semi-automatically on distinctive locations \citep{murphy2011semi}. Two cases (12 and 19) are excluded because of the high uncertainty in the landmarks annotation \citep{staring2014towards}.

In the DIR-Lab-COPDgene database, ten cases with severe breathing disorders are available in inhale and exhale phases. The average image size and the average voxel size are \mbox{$512\times 512 \times 120$} and \mbox{$0.64\times 0.64 \times 2.50$ mm}, \mbox{res}pectively. In each pair, 300 landmarks are annotated.

In the DIR-Lab-4DCT database ten cases are available. We use two phases of the available data: maximum inhalation and maximum exhalation. The size of the images is about $256\times 256 \times 103$ with an average voxel size of \mbox{$1.10\times 1.10 \times 2.50$ mm}. 

Since the convolutional neural networks process the images in a voxel-based manner, all images are resampled to an isotropic voxel size of \mbox{$1.0\times 1.0 \times 1.0$ mm}. 

\subsection{Evaluation measures} 

We use two measures to evaluate the performance of the proposed CNNs: 
\begin{itemize}
\item \textbf{TRE:} The target registration error (TRE) defined as the mean Euclidean distance after registration between corresponding landmarks:
\begin{align}
\mathrm{TRE} = \frac{1}{n} \sum_{i=1}^{n}
\|\bm{T'}(\bm{x}_{Fi}) + \bm{x}_{Fi} - \bm{x}_{Mi} \|_2,
\label{eq:TRE}
\end{align}
where $\bm{x}_F$ and $\bm{x}_M$ are the landmark locations on the fixed and moving images, respectively.     	
\item \textbf{Jac:} The Determinant of the Jacobian of the predicted DVF is calculated in order to measure relative changes in local volume. A very large ($\mathrm{Jac} \gg 1$) or very small ($\mathrm{Jac} \ll 1$) or negative Jac ($ \mathrm{Jac} < 0$) can indicate poor registration quality. We report the percentage of negative Jacobian as well as the standard deviation of the Jacobian inside the lung masks.
\end{itemize}

All of the assessments are performed on the real images.

\subsection{Experimental setup} 

\subsubsection{Training data} \label{sec:NetworkDetails}

In the SPREAD database, 10 patients (20 images) are used for training, 1 patient (2 images) is in the validation set and 8 patients remain for the test set. From the DIR-Lab-COPD database, the first 9 cases (18 images) are used for training, and the remaining case (2 images) is used in the validation set. The entire DIR-Lab-4DCT database is used as an independent test set. The validation set is mainly used for tuning the hyper-parameters and selecting the best network design. In all evaluations, images are multiplied with the lung masks.

To generate training pairs we use the 14 basis types of artificial generations (see Section \ref{sec:extensive_generation}). For each of the three networks, from each original image we generate 70 (5$\times$basis), 42 (3$\times$basis) and 28 (2$\times$basis) artificial pairs in the first stage (RegNet$^4$), the second stage (RegNet$^2$) and the third stage (RegNet$^1$), respectively. Here we generate more images for more coarse stages, as these images are smaller.

In the training phase of the patch-based networks (MV, Uadv), the batch size is 15. The number of patches per pair is 5, 20 and 50 for stage 4, 2 and 1, respectively. The patch size is $101^3$ and $105^3$ for the U-Net-advanced and Multi-view design. 
When choosing samples, several balancing criterion are considered based on the magnitude of DVFs of the patches. An equal number of samples are selected from the range \mbox{[0, 1.5)}, \mbox{[1.5, 8)} and \mbox{[8, 20) mm} for stage 4. For stage 2 and 1 these bins are selected as \mbox{[0, 1.5)}, \mbox{[1.5, 4)}, \mbox{[4, 15) mm} and \mbox{[0, 2)}, \mbox{[2, 7) mm}, respectively. 
Training is run for 30 semi-epochs. All methods are trained with an additional data augmentation step, by adding Gaussian noise to all patches on the fly.

\subsubsection{Software}
In order to efficiently implement the artificial deformation and training phase, we utilize two processes. The task of the first process is to create artificial DVFs and deformed images and write them to disk. The second process has a multithreading paradigm which loads the data from disk and also handles the network training on the GPU.

The CNNs are implemented in Tensorflow.
% \citep{abadi2016tensorflow}.
Artificial DVFs are generated with the help of SimpleITK.
%\citep{lowekamp2013design}.
The code is publicly available via \href{https://github.com/hsokooti/regnet}{github.com/hsokooti/RegNet}.

\subsubsection{\texttt{elastix}}
We compare the proposed CNN-based registration methods with conventional image registration, using \texttt{elastix}.
%\citep{klein2010elastix}. 
We used the following settings: metric: mutual information, optimizer: adaptive stochastic gradient descent, transform: B-spline, number of resolutions: 3, number of iterations per resolution: 500. For the public DIR-Lab-4DCT data, more conventional and CNN-based methods are compared with RegNet in Section \ref{sec:exp:independent_test}.

\subsection{Experiments} 

\subsubsection{Architecture selection}

In order to inspect the performance of the different architectures an evaluation is performed on all pairs in the training and validation sets, i.e. half of the SPREAD data and the entire DIR-Lab-COPDgene data. We utilize the single and mixed category plus identity transform for artificial generations. Please note that the networks are trained with artificial image pairs i.e. during training both the fixed and moving images are deformed versions of the original images. For this evaluation however, we used the original non-deformed pairs, which the network has not seen.

As a first experiment we train and validate the networks on the original image resolution only, i.e. without any multi-stage pipeline: see MV$^1$ and Uadv$^1$ in Section \ref{tb:validation}. It can be seen that the TREs of these networks are on the high end for both databases.
Please note that due to high intensity variation in the baseline and follow-up images in the DIR-Lab-COPDgene database, the overall results are relatively poor. We discuss this issue later in Section \ref{sec::discussion}.

In a second experiment, we train and test the networks on the lowest image resolution only, again without any multi-stage pipeline: see U$^4$, MV$^4$ and Uadv$^4$ in Table \ref{tb:validation}. Note that on the SPREAD data, the performance improved with respect to registration on the original resolution. The main reason is that the lowest resolution training set has the maximum deformation $\theta$ of \mbox{20 mm}, whereas the maximum deformation was set to \mbox{7 mm} in the original resolution training set (see \mbox{Table \ref{tb:DVF_freq}}).
On the DIR-Lab-COPDgene data, similar results were obtained except for U$^4$. 

In the next experiment, we utilized two stages at image resolutions downsampled with a factor of 4 and then 2. In all four tested architectural combinations, the TRE results are better than the single stage networks in both databases, which shows that adding a second stage can improve the performance of RegNet.

Finally, when the original resolution is added to form a three-stage network a small improvement is observed in both databases. By comparing the final TRE results, it can be seen that the performance of all four network combinations are similar. For the remainder of the experiments, we choose the combination (U$^4$-Uadv$^2$-Uadv$^1$) as it obtained slightly better results on the SPREAD database. A Wilcoxon signed-rank test is performed between U$^4$-Uadv$^2$-Uadv$^1$ and other combination in Table \ref{tb:validation}. A statistically significant difference (with $p < 0.05$) between U$^4$-Uadv$^2$-Uadv$^1$ and all single stage and two stages combination can be observed.

\begin{table*}[tb]
\centering \caption{Quantitative results on the training and validation sets. The target registration error (TRE) is reported, together with the percenage of folding and the standard deviation of the Jacobian inside the lung masks. The networks are trained using artificial deformations from the single and mixed category plus identity (see Section \ref{sec:artificial_dvf}). U, Uadv and MV represent the U-Net, U-Net-advanced and Multi-view design (see Section \ref{sec:methods:network}). A Wilcoxon signed-rank test is performed between U$^4$-Uadv$^2$-Uadv$^1$ and others, where $\dagger$ indicates a statistically significant difference with $p < 0.05$.}\label{tb:validation}

%\ccg{SPREAD (case 1-11)} &\ccg{\phantom{x}} &\ccg{\phantom{x}} &\ccg{\phantom{x}} &\ccg{\phantom{x}}&\ccg{\phantom{x}} &\ccg{\phantom{x}}\\
\begin{tabular}{L{2.5cm}lllllll}
\hline
\multirow{2}{*}{Network combination}  &\multicolumn{3}{c}{SPREAD (case 1-11)} & &\multicolumn{3}{c}{DIR-Lab-COPDgene}  \\
\cmidrule(lr){2-4}\cmidrule(lr){6-8}

 & TRE (mm) &\%folding &std(Jac) & & TRE (mm) &\%folding &std(Jac)\\
\hline

MV$^1$ &\scriptsize$3.86{\pm}4.32^\dagger$  &\scriptsize$0.23$\scriptsize${\pm}0.18$ &\scriptsize$0.23$\scriptsize${\pm}0.05$&  &\scriptsize$9.28{\pm}5.83^\dagger$ &\scriptsize$0.24$\scriptsize${\pm}0.07$ &\scriptsize$0.29$\scriptsize${\pm}0.03$\\
Uadv$^1$ &\scriptsize$3.80{\pm}4.15^\dagger$ &\scriptsize$0.24$\scriptsize${\pm}0.20$ & \scriptsize$0.28$\scriptsize${\pm}0.06$ & &\scriptsize$9.65{\pm}6.19^\dagger$ &\scriptsize$0.32$\scriptsize${\pm}0.13$ &\scriptsize$0.35$\scriptsize${\pm}0.05$\\
\phantom{x} &\phantom{x} &\phantom{x} &\phantom{x}\\

U$^4$ &\scriptsize$2.71{\pm}1.59^\dagger$  &\scriptsize$0.00$\scriptsize${\pm}0.00$ &\scriptsize$0.09$\scriptsize${\pm}0.02$ & &\scriptsize$10.2{\pm}6.00^\dagger$  &\scriptsize$0.00$\scriptsize${\pm}0.00$ &\scriptsize$0.10$\scriptsize${\pm}0.01$\\
MV$^4$ &\scriptsize$2.30{\pm}1.80^\dagger$  &\scriptsize$0.00$\scriptsize${\pm}0.00$ &\scriptsize$0.11$\scriptsize${\pm}0.02$ & &\scriptsize$8.27{\pm}5.44^\dagger$  &\scriptsize$0.00$\scriptsize${\pm}0.00$ &\scriptsize$0.14$\scriptsize${\pm}0.02$\\
Uadv$^4$ &\scriptsize$2.29{\pm}1.89^\dagger$ &\scriptsize$0.00$\scriptsize${\pm}0.00$  &\scriptsize$0.08$\scriptsize${\pm}0.01$ & &\scriptsize$8.60{\pm}5.50^\dagger$ &\scriptsize$0.00$\scriptsize${\pm}0.00$ &\scriptsize$0.12$\scriptsize${\pm}0.01$\\
\phantom{x} &\phantom{x} &\phantom{x} &\phantom{x}\\

MV$^4$-MV$^2$ &\scriptsize$1.70{\pm}1.31^\dagger$ &\scriptsize$0.00$\scriptsize${\pm}0.01$ &\scriptsize$0.18$\scriptsize${\pm}0.03$ & &\scriptsize$6.67{\pm}5.53^\dagger$  &\scriptsize$0.01$\scriptsize${\pm}0.01$ &\scriptsize$0.28$\scriptsize${\pm}0.05$\\
U$^4$-MV$^2$ &\scriptsize$1.71{\pm}1.23^\dagger$ &\scriptsize$0.00$\scriptsize${\pm}0.00$ &\scriptsize$0.15$\scriptsize${\pm}0.03$ & &\scriptsize$8.94{\pm}6.95^\dagger$ &\scriptsize$0.03$\scriptsize${\pm}0.02$ &\scriptsize$0.27$\scriptsize${\pm}0.06$\\
Uadv$^4$-Uadv$^2$ &\scriptsize$1.69{\pm}1.34^\dagger$  &\scriptsize$0.00$\scriptsize${\pm}0.00$ &\scriptsize$0.12$\scriptsize${\pm}0.02$ & &\scriptsize$6.96{\pm}5.89^\dagger$ &\scriptsize$0.00$\scriptsize${\pm}0.00$ &\scriptsize$0.21$\scriptsize${\pm}0.04$\\
U$^4$-Uadv$^2$ &\scriptsize$1.68{\pm}1.15^\dagger$ &\scriptsize$0.00$\scriptsize${\pm}0.00$ &\scriptsize$0.11$\scriptsize${\pm}0.02$ & &\scriptsize$8.54{\pm}6.91^\dagger$ &\scriptsize$0.00$\scriptsize${\pm}0.00$ &\scriptsize$0.20$\scriptsize${\pm}0.04$\\
\phantom{x} &\phantom{x} &\phantom{x} &\phantom{x}\\

MV$^4$-MV$^2$-MV$^1$ &\scriptsize$1.63{\pm}1.30^\dagger$ &\scriptsize$0.05$\scriptsize${\pm}0.07$ &\scriptsize$0.22$\scriptsize${\pm}0.04$ & &\scriptsize$6.35{\pm}5.74^\dagger$ &\scriptsize$0.44$\scriptsize${\pm}0.24$ &\scriptsize$0.38$\scriptsize${\pm}0.08$\\
U$^4$-MV$^2$-MV$^1$ &\scriptsize$1.60{\pm}1.20$ &\scriptsize$0.02$\scriptsize${\pm}0.03$ &\scriptsize$0.19$\scriptsize${\pm}0.04$ & &\scriptsize$8.65{\pm}7.27^\dagger$ &\scriptsize$0.49$\scriptsize${\pm}0.27$ &\scriptsize$0.38$\scriptsize${\pm}0.09$\\
Uadv$^4$-Uadv$^2$-Uadv$^1$ &\scriptsize$1.60{\pm}1.23$ &\scriptsize$0.03$\scriptsize${\pm}0.04$ &\scriptsize$0.22$\scriptsize${\pm}0.03$ & &\scriptsize$6.45{\pm}6.39^\dagger$ &\scriptsize$0.29$\scriptsize${\pm}0.20$ &\scriptsize$0.37$\scriptsize${\pm}0.10$\\
U$^4$-Uadv$^2$-Uadv$^1$ &\scriptsize$1.57{\pm}1.15$ &\scriptsize$0.02$\scriptsize${\pm}0.02$ &\scriptsize$0.20$\scriptsize${\pm}0.03$ & &\scriptsize$8.07{\pm}7.65$ &\scriptsize$0.39$\scriptsize${\pm}0.24$ &\scriptsize$0.38$\scriptsize${\pm}0.10$\\
\hline
\end{tabular}
\end{table*}

\subsubsection{Independent test set experiments} \label{sec:exp:independent_test}

Now that we have selected the best network combination, we applied the U$^4$-Uadv$^2$-Uadv$^1$ pipeline on the independent test set (without retraining): 8 cases of the SPREAD database and the complete DIR-Lab-4DCT database. The results are given in Tables \ref{tb:Test-SPREAD} and \ref{tb:Test-DIR-Lab-4DCT}. 

For the SPREAD database, the TRE results with affine and B-spline registration are compared with three versions of RegNet trained using the category ``S'' (single frequency plus identity), ``S+M'' (single frequency and mixed frequency plus identity) and ``S+M+R'' (single frequency plus mixed frequency and respiratory motion plus identity). 
Since there is no respiratory motion in the SPREAD data, adding respiratory motion did not improve the performance of the registration. 
Adding mixed frequencies did not change the results considerably: there was a small improvement for the cases 1-11, and slightly larger TREs for the cases 13 to 21. The percentage of folding inside the lung masks for the RegNet trained using ``S'' is also available in Table \ref{tb:Test-SPREAD}, which reports that the percentage of negative Jacobian are small in most cases, especially, when the TRE after affine registration is not very large.
A Wilcoxon signed-rank test is performed between the \texttt{elastix} B-spline and other results. It can be seen that in most cases there is no significant difference between B-spline registration and RegNet trained using ``S'' or trained using ``S+M''.

\begin{figure*}[tb]
\begin{subfigure}{0.16\textwidth}
	\centering
	\caption{Fixed}
	\includegraphics[width=1\textwidth, trim={150 250 80 200}, clip]{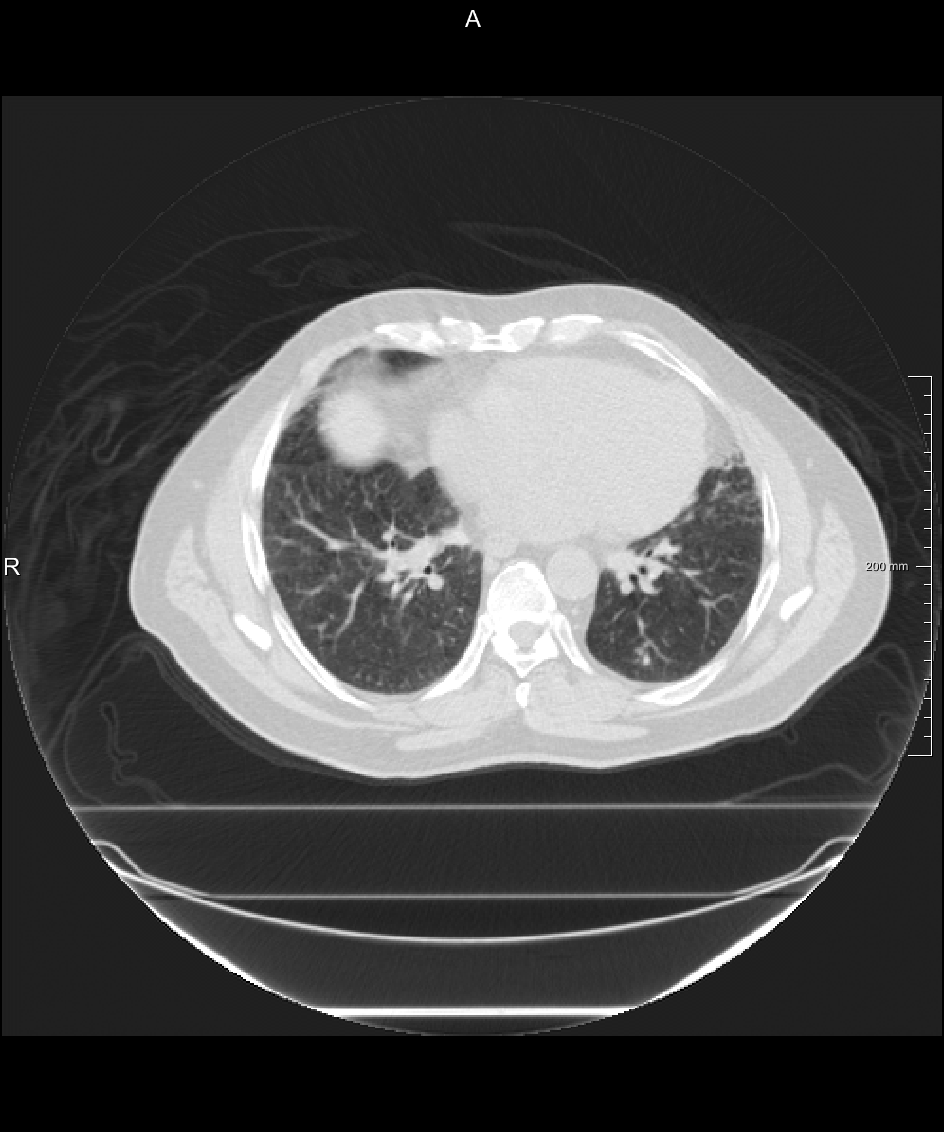}				
\end{subfigure}
\begin{subfigure}{0.16\textwidth}
	\centering
	\caption{Affine}
	\includegraphics[width=1\textwidth, trim={150 250 80 200}, clip]{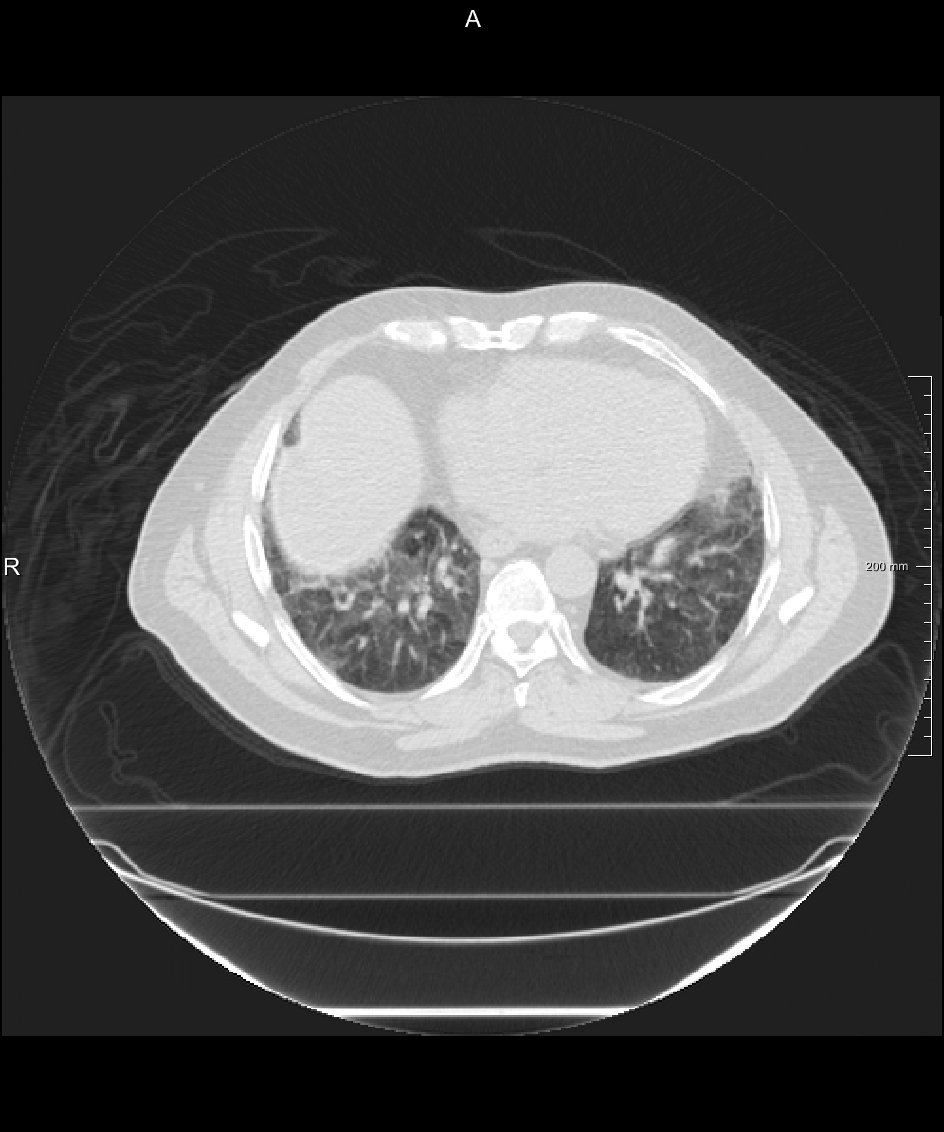}				
\end{subfigure}
\begin{subfigure}{0.16\textwidth}
	\centering
	\caption{B-spline}
	\includegraphics[width=1\textwidth, trim={150 250 80 200}, clip]{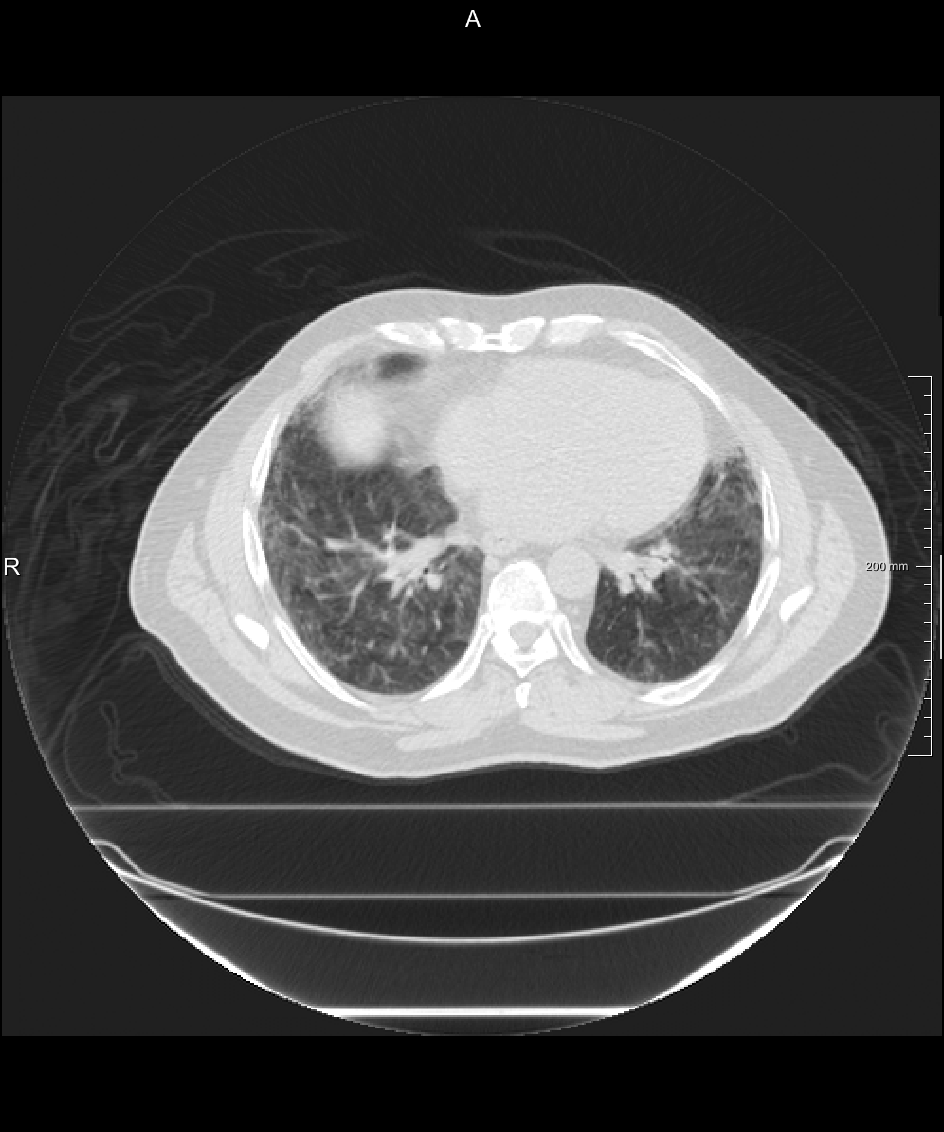}				
\end{subfigure}
\begin{subfigure}{0.16\textwidth}
	\centering
	\caption{RegNet ``S''}
	\includegraphics[width=1\textwidth, trim={150 250 80 200}, clip]{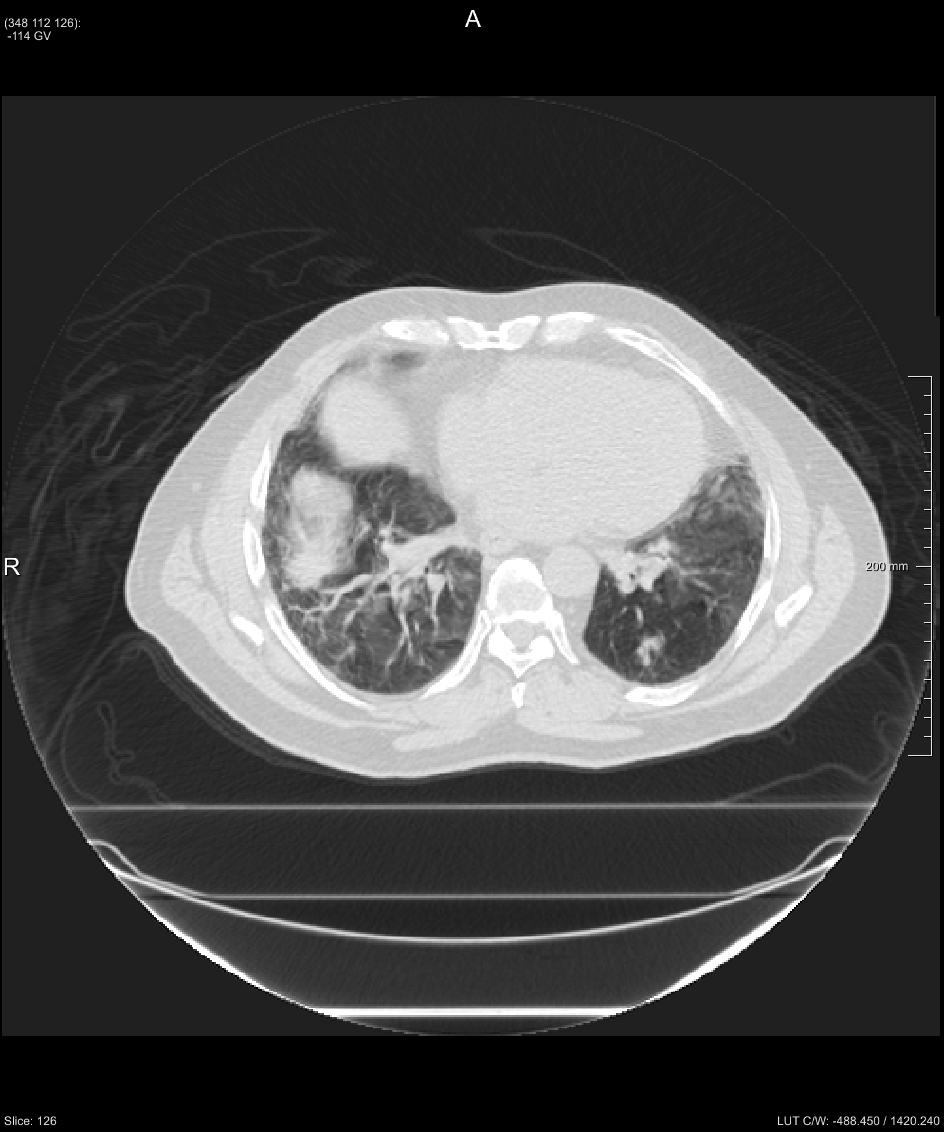}				
\end{subfigure}
\begin{subfigure}{0.16\textwidth}
	\centering
	\caption{RegNet ``S+M''}
	\includegraphics[width=1\textwidth, trim={150 250 80 200}, clip]{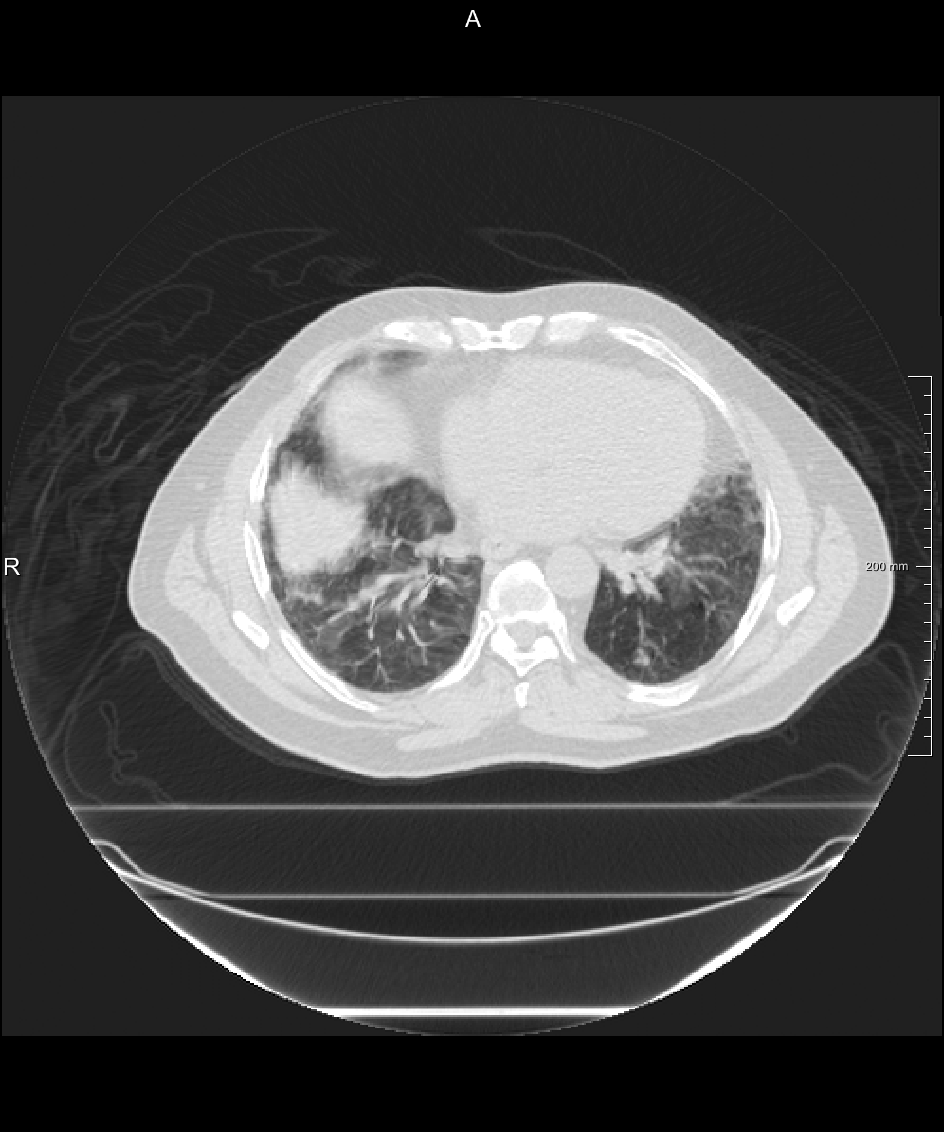}				
\end{subfigure}
\begin{subfigure}{0.16\textwidth}
	\centering
	\caption{RegNet ``S+M+R''}
	\includegraphics[width=1\textwidth, trim={150 250 80 200}, clip]{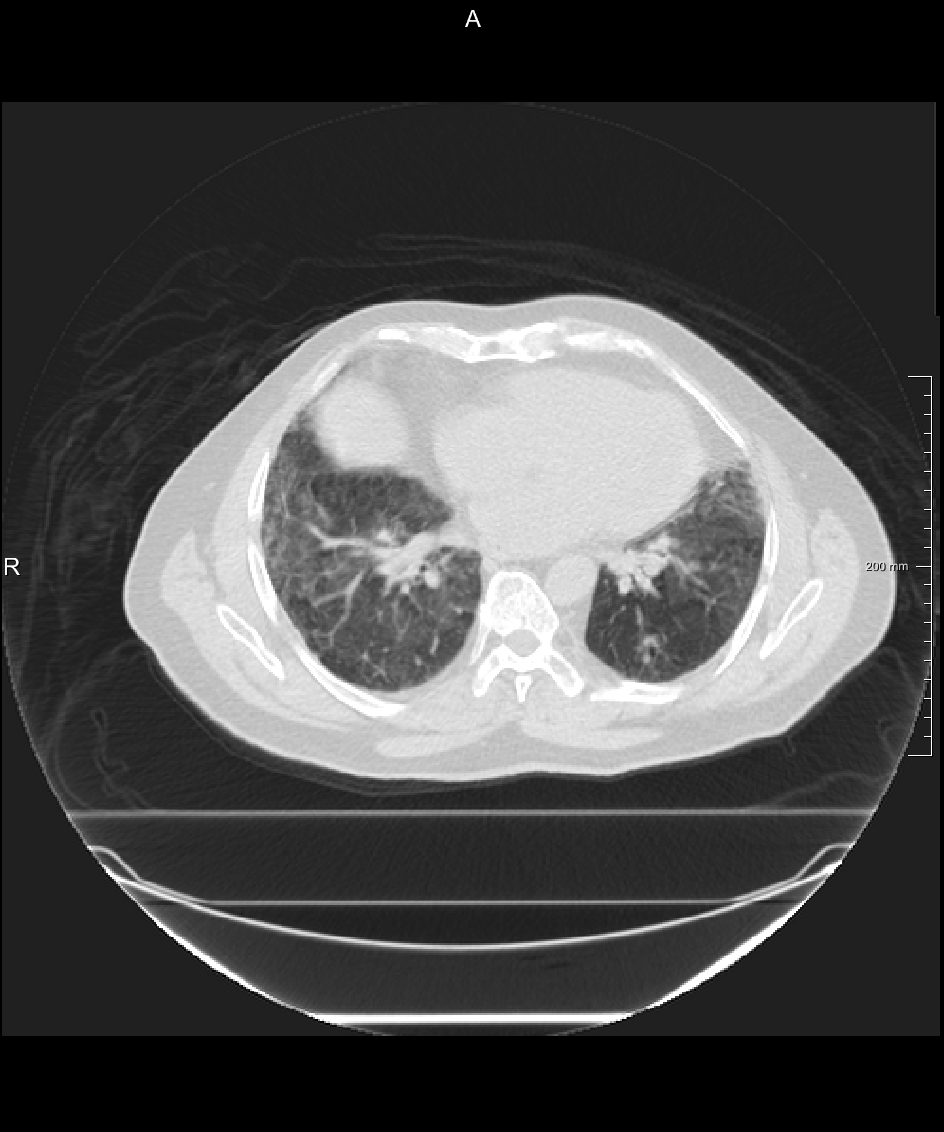}				
\end{subfigure}
\vspace{0.11cm}
\\
\phantom{\begin{subfigure}{0.16\textwidth}
	\centering
	\includegraphics[width=1\textwidth, trim={150 250 80 200}, clip]
	{IM_Fixed.png}				
\end{subfigure}}
\begin{subfigure}{0.16\textwidth}
	\centering
	\includegraphics[width=1\textwidth, trim={150 250 80 200}, clip]{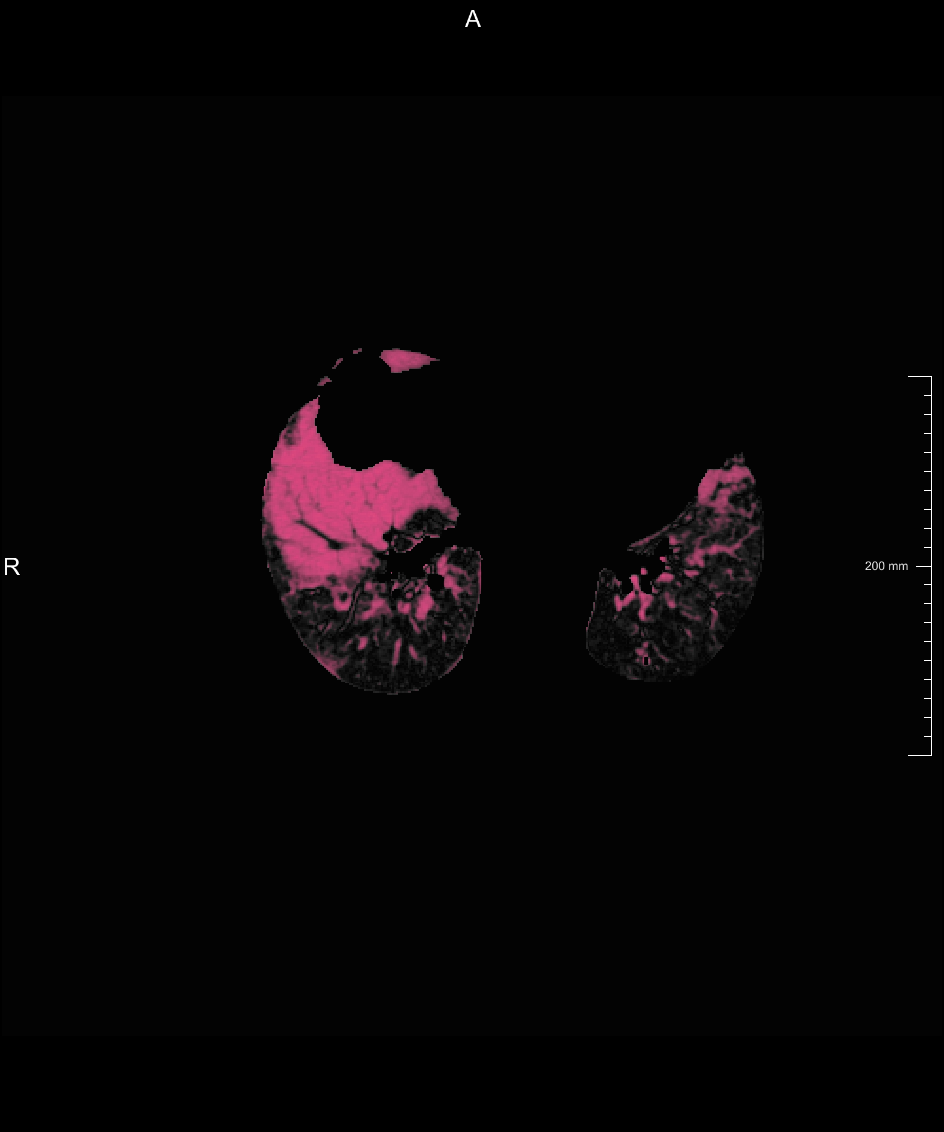}				
\end{subfigure}
\begin{subfigure}{0.16\textwidth}
	\centering
	\includegraphics[width=1\textwidth, trim={150 250 80 200}, clip]{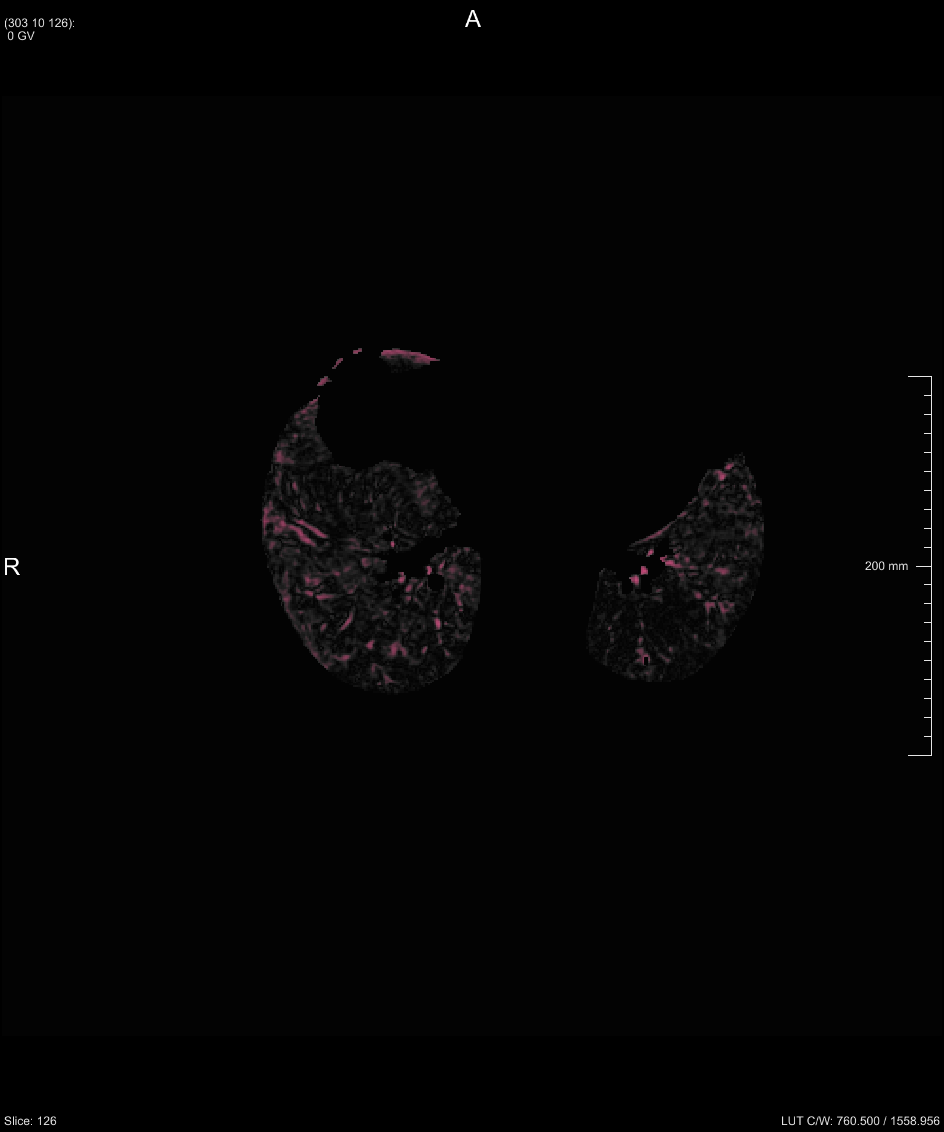}				
\end{subfigure}
\begin{subfigure}{0.16\textwidth}
	\centering
	\includegraphics[width=1\textwidth, trim={150 250 80 200}, clip]{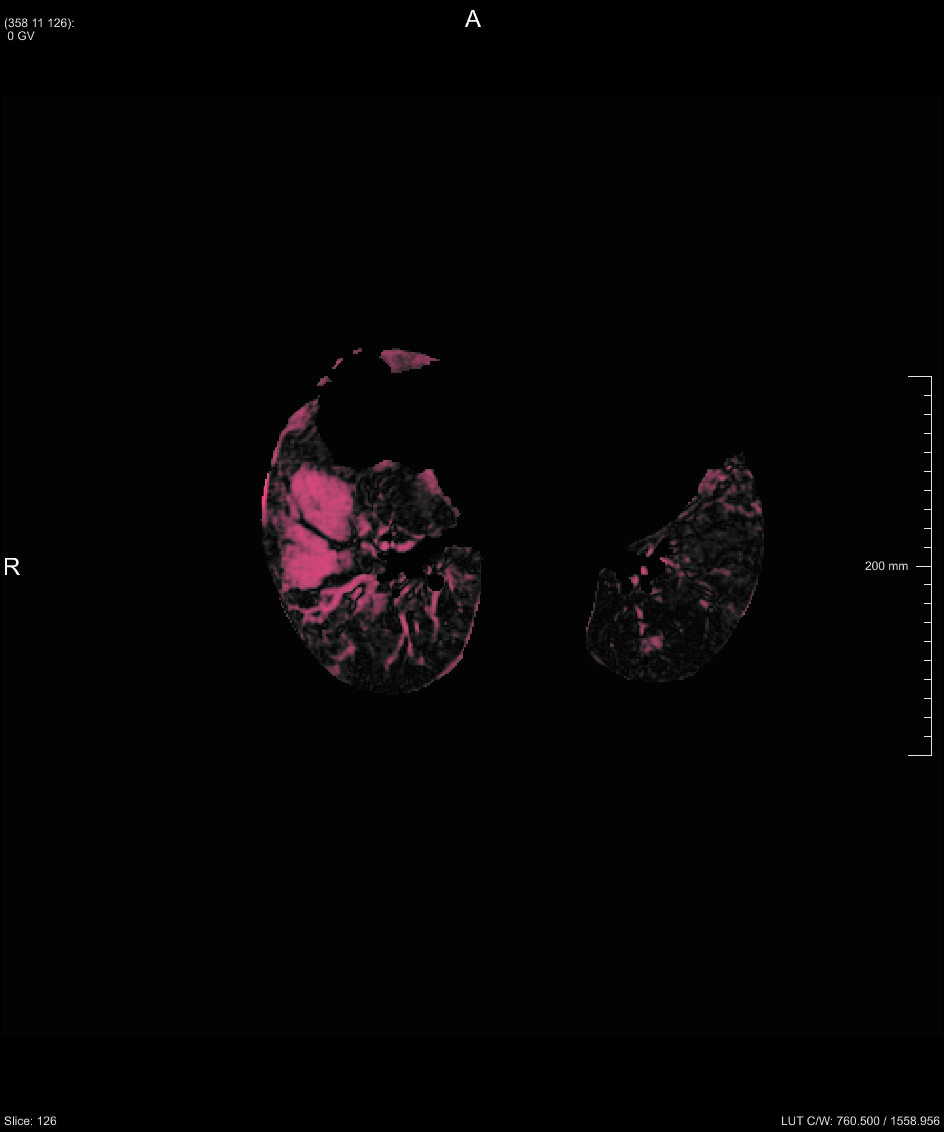}				
\end{subfigure}
\begin{subfigure}{0.16\textwidth}
	\centering
	\includegraphics[width=1\textwidth, trim={150 250 80 200}, clip]{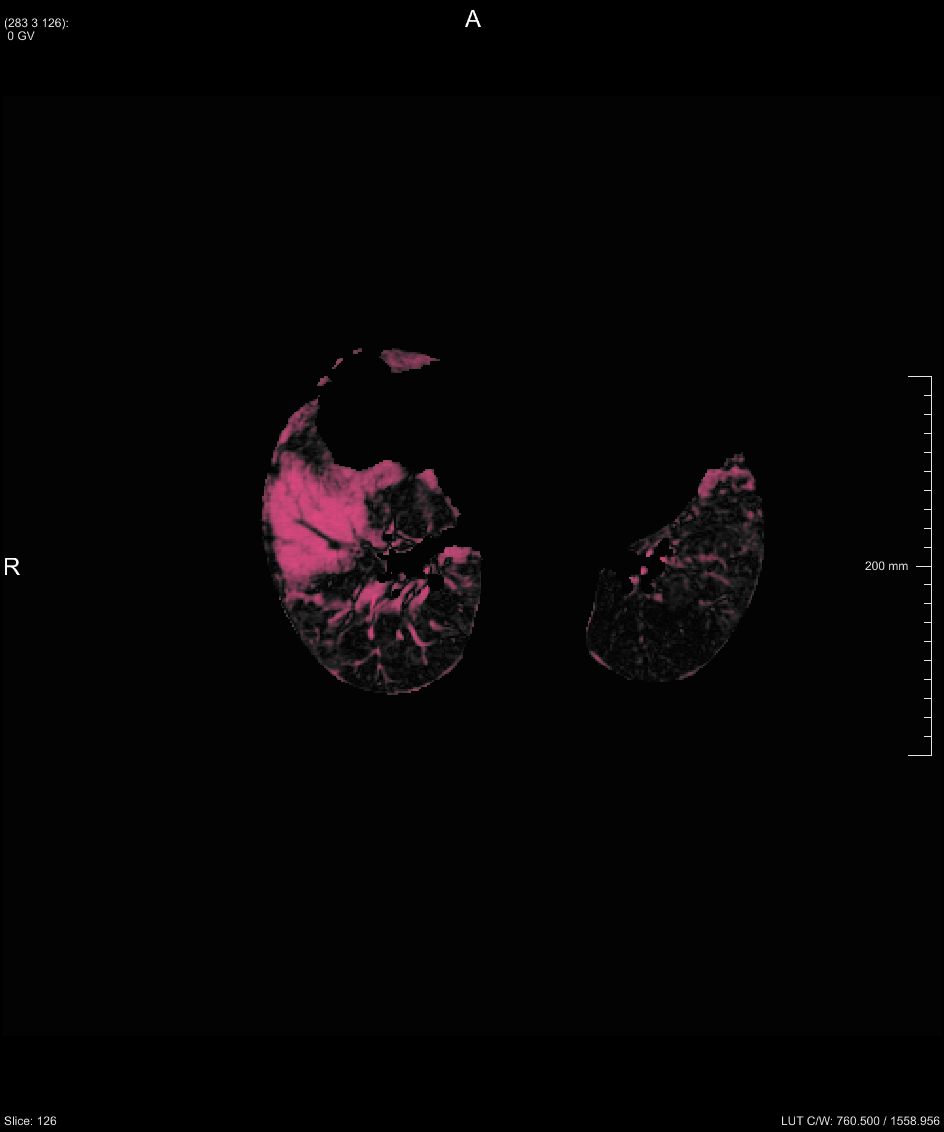}				
\end{subfigure}
\begin{subfigure}{0.16\textwidth}
	\centering
	\includegraphics[width=1\textwidth, trim={150 250 80 200}, clip]{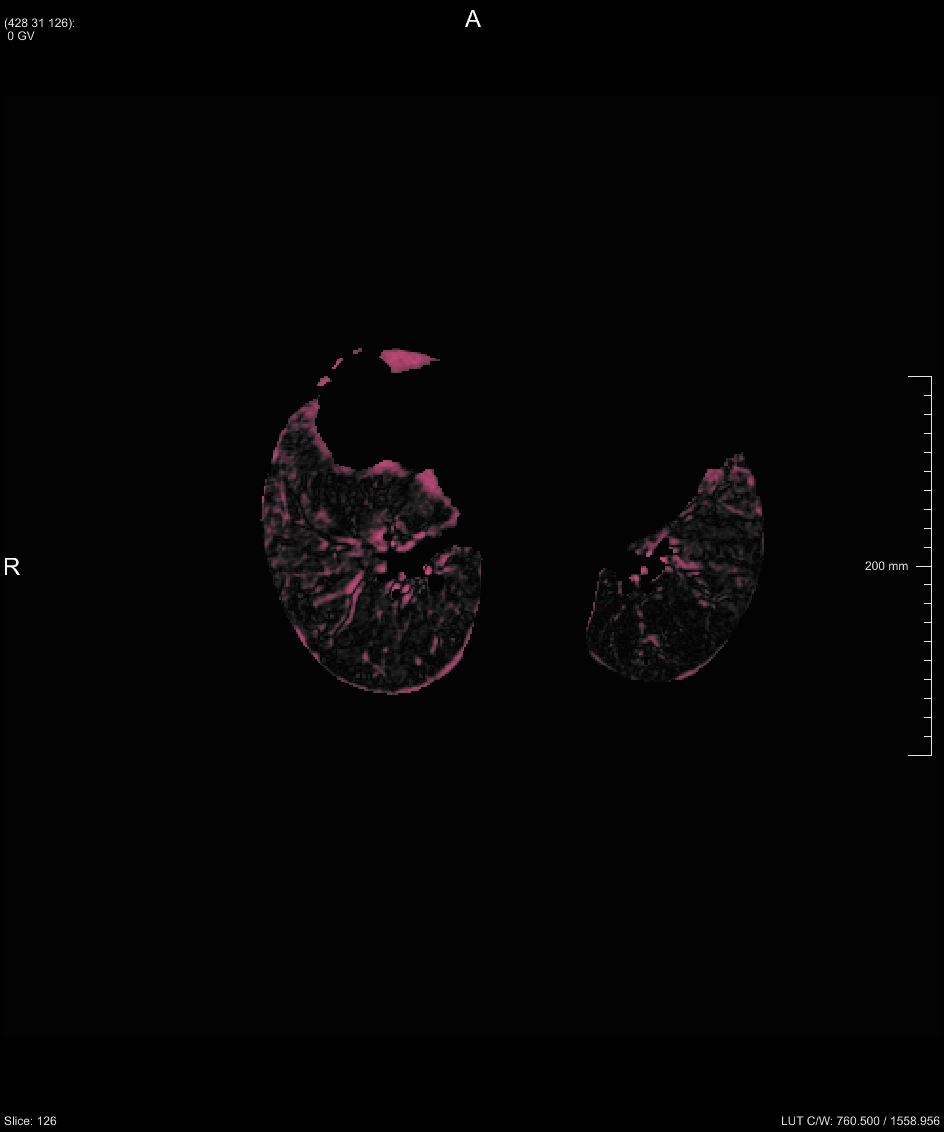}				
\end{subfigure}
\caption{Example results (top row) and difference images (bottom
row) from DIR-Lab-4DCT study.} \label{fig:results_quantitative}
\end{figure*}

% uncomment
For the DIR-Lab-4DCT database, a comparison between RegNet and affine, B-spline (three resolutions), an advanced conventional registration method using sliding motion \citep{berendsen2014registration} and three other CNN-based methods \citep{eppenhof2018pulmonary, de2019deep, sentker2018gdl} is available in Table \ref{tb:Test-DIR-Lab-4DCT}.
It can be seen that training with ``S+M'' improved performance slightly with respect to just ``S''. Adding the respiratory motion category improved performance substantially, as these are inhale-exhale pairs; this is predominantly caused by the patients where the TRE after affine registration was still quite large. An example visualization is also available in Fig. \ref{fig:results_quantitative}, showing that adding the respiratory motion category can align images better in the diaphragm region.
The advanced conventional registration method that leverages sliding motion \citep{berendsen2014registration} is still better than RegNet.
Note that RegNet was not trained on the DIR-Lab-4DCT data, similar to
\cite{eppenhof2018pulmonary, sentker2018gdl}. However, \cite{de2019deep} and \cite{eppenhof2018pulmonary}-DIR methods were trained on the same database but using cross-validation to report the results.
Also note that the results reported \mbox{in \cite{sentker2018gdl}} are averaged over all phases of DIR-Lab-4DCT (T00 to T10), while the results of other CNN methods (including RegNet) are reported between the maximum inhale and maximum exhale phase (T00 and T50). These reported results are therefore likely somewhat better than the results for T00 and T50 only.
% uncomment

%
%\hl{add this paragraph}
%For the DIR-Lab-4DCT database, the settings are taken from \cite{berendsen2014registration}: metric: \pdfmarkupcomment[markup=Highlight,color=CommentColor]{?}{Floris' parameter file is not available in the elastix website. Do you have access to the parameter file?}, optimizer: gradient descent, transform: B-spline, number of resolutions: 6, number of iterations per resolution: 2000. 

\subsubsection{Inference}
At inference time, the patch size can be enlarged depending on the available GPU memory. For the U-Net-advanced design, the inference time of an image of size $101^3$ and $269^3$ voxels, is 0.02 s and 2.4 s, respectively, on our TITAN Xp (12 GB). An image of size $273^3$ voxels took about 2.1 s to process for the Multi-view design. For the U-Net design we used the downsized image (by a factor of 4) of size $125^3$ which took 0.02 s to be processed.

\begin{table*}[tb]
\centering \caption{Quantitative results of the SPREAD study in the training set (case 1 to case 11) and in the test set (case 13 to case 21). This experiment is performed with the network combination U$^4$-Uadv$^2$-Uadv$^1$. The target registration error (TRE) is reported, together with the percentage of folding and the standard deviation of the Jacobian inside the lung masks. S, M and R indicate single frequency, mixed frequency and respiratory motion, respectively (see Section \ref{sec:artificial_dvf}). A Wilcoxon signed-rank test is performed between the B-spline registration and others. The symbol $\dagger$ indicates a significant difference between the average of TRE of B-spline registration and others, where $\dagger$ indicates a statistically significant difference with $p < 0.05$. The best method is shown in bold and the second best method is shown in green.} \label{tb:Test-SPREAD}

\begin{tabular}{lllllL{1.55cm}L{1.55cm}L{1.55cm}L{1.4cm}L{1.4cm}}
\hline
 &\multirow{1}{*}{\texttt{elastix}} & \multicolumn{3}{c}{\texttt{elastix} B-spline} &\multicolumn{5}{c}{RegNet} \\
\cmidrule(lr){3-5} \cmidrule(lr){6-10} 
 &Affine &\multicolumn{3}{c}{} &S &S+M &S+M+R &S & S\\

pair &TRE (mm) &TRE (mm) &\%folding &std(Jac) &TRE (mm) &TRE (mm) &TRE (mm) &\%folding &std(Jac) \\

\hline

case 1 &\scriptsize$8.77{\pm}2.76^\dagger$ &\scriptsize$2.13{\pm}2.12$  &\scriptsize$0.00$ &\scriptsize$0.19$&\scriptsize\tcb{$1.87{\pm}1.68$} &\scriptsize\boldmath{$1.78{\pm}1.66$} &\scriptsize$1.92{\pm}1.71$ &\scriptsize$0.20$ &\scriptsize$0.26$\\
case 2 &\scriptsize$7.41{\pm}2.99^\dagger$ &\scriptsize$1.48{\pm}1.18$  &\scriptsize$0.00$ &\scriptsize$0.11$&\scriptsize\tcb{$1.44{\pm}0.92$} &\scriptsize\boldmath{$1.37{\pm}0.88$} &\scriptsize$1.54{\pm}1.03$ &\scriptsize$0.03$ &\scriptsize$0.20$\\
case 3 &\scriptsize$4.34{\pm}1.89^\dagger$ &\scriptsize\boldmath{$1.66{\pm}1.13$}  &\scriptsize$0.00$ &\scriptsize$0.09$&\scriptsize\tcb{$1.78{\pm}1.18$} &\scriptsize$1.79{\pm}1.21$ &\scriptsize$1.81{\pm}1.19^\dagger$ &\scriptsize$0.00$ &\scriptsize$0.15$\\
case 4 &\scriptsize$11.4{\pm}3.44^\dagger$ &\scriptsize$1.79{\pm}1.43$  &\scriptsize$0.00$ &\scriptsize$0.15$&\scriptsize\boldmath{$1.70{\pm}1.41$} &\scriptsize\tcb{$1.70{\pm}1.29$} &\scriptsize$1.99{\pm}1.74$ &\scriptsize$0.07$ &\scriptsize$0.21$\\
case 5 &\scriptsize$6.47{\pm}2.07^\dagger$ &\scriptsize\boldmath{$1.08{\pm}0.62$}  &\scriptsize$0.00$ &\scriptsize$0.09$&\scriptsize\tcb{$1.15{\pm}0.76$} &\scriptsize$1.17{\pm}0.70$ &\scriptsize$1.24{\pm}0.78^\dagger$ &\scriptsize$0.00$ &\scriptsize$0.15$\\
case 6 &\scriptsize$8.22{\pm}2.37^\dagger$ &\scriptsize$2.06{\pm}1.37$  &\scriptsize$0.00$ &\scriptsize$0.14$&\scriptsize$1.98{\pm}1.44$ &\scriptsize\boldmath{$1.90{\pm}1.27$} &\scriptsize\tcb{$1.92{\pm}1.22$} &\scriptsize$0.02$ &\scriptsize$0.21$\\
case 7 &\scriptsize$5.51{\pm}1.38^\dagger$ &\scriptsize\boldmath{$1.50{\pm}1.11$}  &\scriptsize$0.00$ &\scriptsize$0.10$&\scriptsize$1.70{\pm}1.07^\dagger$ &\scriptsize$1.63{\pm}1.22$ &\scriptsize\tcb{$1.51{\pm}1.10$} &\scriptsize$0.00$ &\scriptsize$0.17$\\
case 8 &\scriptsize$3.67{\pm}2.31^\dagger$ &\scriptsize$1.70{\pm}1.23$  &\scriptsize$0.00$ &\scriptsize$0.14$&\scriptsize$1.74{\pm}1.02$ &\scriptsize\tcb{$1.63{\pm}0.94$} &\scriptsize\boldmath{$1.53{\pm}0.84$} &\scriptsize$0.01$ &\scriptsize$0.20$\\
case 9 &\scriptsize$4.93{\pm}1.61^\dagger$ &\scriptsize\boldmath{$1.28{\pm}0.72$}  &\scriptsize$0.00$ &\scriptsize$0.09$&\scriptsize\tcb{$1.35{\pm}0.72$} &\scriptsize$1.41{\pm}0.87$ &\scriptsize$1.51{\pm}0.77^\dagger$ &\scriptsize$0.02$ &\scriptsize$0.16$\\
case 10 &\scriptsize$6.22{\pm}2.27^\dagger$ &\scriptsize\boldmath{$1.33{\pm}1.11$}  &\scriptsize$0.00$ &\scriptsize$0.10$&\scriptsize$1.40{\pm}0.90$ &\scriptsize\tcb{$1.40{\pm}0.87$} &\scriptsize$1.43{\pm}0.85^\dagger$ &\scriptsize$0.02$ &\scriptsize$0.18$\\
case 11 &\scriptsize$5.93{\pm}2.20^\dagger$ &\scriptsize\boldmath{$1.40{\pm}1.10$}  &\scriptsize$0.00$ &\scriptsize$0.13$&\scriptsize$1.49{\pm}1.15$ &\scriptsize\tcb{$1.44{\pm}1.19$} &\scriptsize$1.51{\pm}1.08$ &\scriptsize$0.03$ &\scriptsize$0.21$\\
\rule{0pt}{3ex}Total &\scriptsize$6.62{\pm}3.17^\dagger$ &\scriptsize\tcb{$1.58{\pm}1.29$} &\scriptsize$0.00$\scriptsize${\pm}0.00$ &\scriptsize$0.12$\scriptsize${\pm}0.03$ &\scriptsize$1.60{\pm}1.17^\dagger$ &\scriptsize\boldmath{$1.57{\pm}1.15$} &\scriptsize$1.63{\pm}1.19^\dagger$ &\scriptsize$0.04$\scriptsize${\pm}0.05$ &\scriptsize$0.19$\scriptsize${\pm}0.03$ \\

\hline

\rule{0pt}{3ex}case 13 &\scriptsize$12.5{\pm}15.8^\dagger$ &\scriptsize$7.94{\pm}16.0$  &\scriptsize$0.00$ &\scriptsize$0.13$&\scriptsize\boldmath{$7.37{\pm}14.0$} &\scriptsize\tcb{$7.76{\pm}15.2$} &\scriptsize$8.28{\pm}15.4$ &\scriptsize$0.71$ &\scriptsize$0.36$\\
case 14 &\scriptsize$8.99{\pm}2.40^\dagger$ &\scriptsize\tcb{$1.86{\pm}1.19$}  &\scriptsize$0.00$ &\scriptsize$0.11$&\scriptsize\boldmath{$1.71{\pm}1.18$} &\scriptsize$2.08{\pm}1.81$ &\scriptsize$2.25{\pm}1.76^\dagger$ &\scriptsize$0.06$ &\scriptsize$0.25$\\
case 15 &\scriptsize$3.17{\pm}1.32^\dagger$ &\scriptsize\boldmath{$1.20{\pm}0.82$}  &\scriptsize$0.00$ &\scriptsize$0.11$&\scriptsize$1.39{\pm}0.86$ &\scriptsize\tcb{$1.29{\pm}0.82$} &\scriptsize$1.33{\pm}0.84$ &\scriptsize$0.00$ &\scriptsize$0.18$\\
case 16 &\scriptsize$8.94{\pm}1.84^\dagger$ &\scriptsize\boldmath{$1.30{\pm}0.80$}  &\scriptsize$0.00$ &\scriptsize$0.09$&\scriptsize\tcb{$1.54{\pm}0.96$} &\scriptsize$1.78{\pm}0.98^\dagger$ &\scriptsize$1.96{\pm}1.01^\dagger$ &\scriptsize$0.00$ &\scriptsize$0.19$\\
case 17 &\scriptsize$13.4{\pm}4.73^\dagger$ &\scriptsize\boldmath{$1.76{\pm}0.73$}  &\scriptsize$0.00$ &\scriptsize$0.09$&\scriptsize$2.89{\pm}3.66^\dagger$ &\scriptsize\tcb{$2.30{\pm}1.70^\dagger$} &\scriptsize$3.37{\pm}3.43^\dagger$ &\scriptsize$0.38$ &\scriptsize$0.27$\\
case 18 &\scriptsize$7.85{\pm}2.89^\dagger$ &\scriptsize$1.65{\pm}1.41$  &\scriptsize$0.00$ &\scriptsize$0.15$&\scriptsize\boldmath{$1.40{\pm}0.86$} &\scriptsize\tcb{$1.60{\pm}1.16$} &\scriptsize$1.71{\pm}1.04$ &\scriptsize$0.02$ &\scriptsize$0.21$\\
case 20 &\scriptsize$4.43{\pm}2.14^\dagger$ &\scriptsize\boldmath{$1.31{\pm}0.90$}  &\scriptsize$0.00$ &\scriptsize$0.11$&\scriptsize\tcb{$1.41{\pm}1.00$} &\scriptsize$1.50{\pm}1.05^\dagger$ &\scriptsize$1.52{\pm}0.97^\dagger$ &\scriptsize$0.14$ &\scriptsize$0.22$\\
case 21 &\scriptsize$6.48{\pm}2.03^\dagger$ &\scriptsize\boldmath{$1.26{\pm}1.35$}  &\scriptsize$0.00$ &\scriptsize$0.09$&\scriptsize$1.36{\pm}1.19$ &\scriptsize\tcb{$1.33{\pm}1.36$} &\scriptsize$1.36{\pm}1.47^\dagger$ &\scriptsize$0.01$ &\scriptsize$0.17$\\
\rule{0pt}{3ex} Total &\scriptsize$8.16{\pm}6.76^\dagger$ &\scriptsize\boldmath{$2.21{\pm}5.86$} &\scriptsize$0.00$\scriptsize${\pm}0.00$ &\scriptsize$0.11$\scriptsize${\pm}0.02$ &\scriptsize\tcb{$2.32{\pm}5.33^\dagger$} &\scriptsize$2.39{\pm}5.64^\dagger$ &\scriptsize$2.65{\pm}5.82^\dagger$ &\scriptsize$0.19$\scriptsize${\pm}0.24$ &\scriptsize$0.24$\scriptsize${\pm}0.06$ \\
\hline

\end{tabular}
\end{table*} 

\newcommand\cwidth{1.4}

\begin{table*}[tb]
\centering \caption{Quantitative results on the DIR-Lab-4DCT study. This experiment is performed with the network combination of U$^4$-Uadv$^2$-Uadv$^1$. The target registration error (TRE) is reported, together with the percentage of folding and the standard deviation of the Jacobian inside the lung masks. S, M and R indicate single frequency, mixed frequency and respiratory motion, respectively (see Section \ref{sec:artificial_dvf}). The result of \cite{sentker2018gdl} is the average of all respiratory phases per each case. The best method is shown in bold and the second best method is shown in green.} \label{tb:Test-DIR-Lab-4DCT}
\resizebox{\textwidth}{!}{
\begin{tabular}{llL{\cwidth cm}L{\cwidth cm}L{\cwidth cm}L{\cwidth cm}L{\cwidth cm}L{\cwidth cm}L{\cwidth cm}L{\cwidth cm}L{\cwidth cm}L{1.3 cm}L{1.3 cm}}
\hline

 &\multirow{3}{*}{\thead{\texttt{elastix} \\Affine}} &\multirow{3}{*}{\thead{\texttt{elastix} \\B-spline}} &\multirow{3}{\cwidth cm}{\cite{berendsen2014registration}} &\multirow{3}{\cwidth cm}{\cite{sentker2018gdl}} &\multirow{3}{\cwidth cm}{\cite{de2019deep}} &\multirow{3}{\cwidth cm}{\cite{eppenhof2018pulmonary}} &\multirow{3}{\cwidth cm}{\cite{eppenhof2018pulmonary}-DIR} & \multicolumn{5}{c}{\multirow{2}{*}{RegNet}} \\
& & & & & & & &  & & & &\\
\cline{9-13}
& & & & & & & & S &S+M &S+M+R &S+M+R &S+M+R\\
pair &TRE (mm) &TRE (mm) &TRE (mm) &TRE (mm) &TRE (mm) &TRE (mm) &TRE (mm) &TRE (mm) &TRE (mm) &TRE (mm) & \%folding &std(Jac) \\
\hline

case 01 &\scriptsize$3.02{\pm}2.13$ &\scriptsize$1.21{\pm}0.71$ &\scriptsize\boldmath{$1.00{\pm}0.52$} &\scriptsize$1.20{\pm}0.60$ &\scriptsize$1.27{\pm}1.16$ &\scriptsize$1.45{\pm}1.06$ &- &\scriptsize\tcb{$1.09{\pm}0.51$} &\scriptsize$1.12{\pm}0.54$ &\scriptsize$1.13{\pm}0.51$ &\scriptsize$0.00$ &\scriptsize$0.10$\\

case 02 &\scriptsize$3.76{\pm}3.20$ &\scriptsize$1.39{\pm}1.27$ &\scriptsize\boldmath{$1.02{\pm}0.57$} &\scriptsize$1.19{\pm}0.63$ &\scriptsize$1.20{\pm}1.12$ &\scriptsize$1.46{\pm}0.76$ &\scriptsize$1.24{\pm}0.61$ &\scriptsize$1.08{\pm}0.89$ &\scriptsize\tcb{$1.06{\pm}0.57$} &\scriptsize$1.08{\pm}0.55$ &\scriptsize$0.00$ &\scriptsize$0.12$\\

case 03 &\scriptsize$5.92{\pm}3.64$ &\scriptsize$2.44{\pm}2.11$ &\scriptsize\boldmath{$1.14{\pm}0.89$} &\scriptsize$1.67{\pm}0.90$ &\scriptsize$1.48{\pm}1.26$ &\scriptsize$1.57{\pm}1.10$ &- &\scriptsize\tcb{$1.23{\pm}0.69$} &\scriptsize$1.23{\pm}0.75$ &\scriptsize$1.33{\pm}0.73$ &\scriptsize$0.00$ &\scriptsize$0.14$\\

case 04 &\scriptsize$9.01{\pm}4.53$ &\scriptsize$2.16{\pm}2.16$ &\scriptsize\boldmath{$1.46{\pm}0.96$} &\scriptsize$2.53{\pm}2.01$ &\scriptsize$2.09{\pm}1.93$ &\scriptsize$1.95{\pm}1.32$ &\scriptsize$1.70{\pm}1.00$ &\scriptsize\tcb{$1.47{\pm}0.95$} &\scriptsize$1.62{\pm}1.09$ &\scriptsize$1.57{\pm}0.99$ &\scriptsize$0.00$ &\scriptsize$0.18$\\

case 05 &\scriptsize$3.95{\pm}2.85$ &\scriptsize$3.02{\pm}3.22$ &\scriptsize$1.61{\pm}1.48$ &\scriptsize$2.06{\pm}1.56$ &\scriptsize$1.95{\pm}2.10$ &\scriptsize$2.07{\pm}1.59$ &- &\scriptsize\boldmath{$1.58{\pm}1.33$} &\scriptsize\tcb{$1.60{\pm}1.33$} &\scriptsize$1.62{\pm}1.30$ &\scriptsize$0.00$ &\scriptsize$0.14$\\

case 06 &\scriptsize$10.7{\pm}6.80$ &\scriptsize$3.33{\pm}3.30$ &\scriptsize\boldmath{$1.42{\pm}1.71$} &\scriptsize$2.90{\pm}1.70$ &\scriptsize$5.16{\pm}7.09$ &\scriptsize$3.04{\pm}2.73$ &- &\scriptsize$4.56{\pm}7.06$ &\scriptsize$4.95{\pm}6.91$ &\scriptsize\tcb{$2.75{\pm}2.91$} &\scriptsize$0.03$ &\scriptsize$0.25$\\

case 07 &\scriptsize$11.1{\pm}7.43$ &\scriptsize$6.16{\pm}6.33$ &\scriptsize\boldmath{$1.49{\pm}4.25$} &\scriptsize$3.60{\pm}2.99$ &\scriptsize$3.05{\pm}3.01$ &\scriptsize$3.41{\pm}2.75$ &- &\scriptsize$6.10{\pm}7.10$ &\scriptsize$5.00{\pm}6.35$ &\scriptsize\tcb{$2.34{\pm}2.32$} &\scriptsize$0.03$ &\scriptsize$0.24$\\

case 08 &\scriptsize$12.0{\pm}6.59$ &\scriptsize$9.36{\pm}9.30$ &\scriptsize\boldmath{$1.62{\pm}1.71$} &\scriptsize$5.29{\pm}5.52$ &\scriptsize$6.48{\pm}5.37$ &\scriptsize\tcb{$2.80{\pm}2.46$} &- &\scriptsize$6.54{\pm}8.51$ &\scriptsize$6.18{\pm}7.01$ &\scriptsize$3.29{\pm}4.32$ &\scriptsize$0.01$ &\scriptsize$0.22$\\

case 09 &\scriptsize$7.89{\pm}3.83$ &\scriptsize$3.31{\pm}2.74$ &\scriptsize\boldmath{$1.30{\pm}0.76$} &\scriptsize$2.38{\pm}1.46$ &\scriptsize$2.10{\pm}1.66$ &\scriptsize$2.18{\pm}1.24$ &\scriptsize\tcb{$1.61{\pm}0.82$} &\scriptsize$2.02{\pm}2.25$ &\scriptsize$1.84{\pm}1.93$ &\scriptsize$1.86{\pm}1.47$ &\scriptsize$0.00$ &\scriptsize$0.17$\\

case 10 &\scriptsize$6.87{\pm}6.12$ &\scriptsize$2.72{\pm}3.43$ &\scriptsize\boldmath{$1.50{\pm}1.31$} &\scriptsize$2.13{\pm}1.88$ &\scriptsize$2.09{\pm}2.24$ &\scriptsize$1.83{\pm}1.36$ &- &\scriptsize$2.82{\pm}4.93$ &\scriptsize$2.44{\pm}3.85$ &\scriptsize\tcb{$1.63{\pm}1.29$} &\scriptsize$0.00$ &\scriptsize$0.19$\\

\rule{0pt}{3ex}Total &\scriptsize$7.43{\pm}5.92$ &\scriptsize$3.51{\pm}4.83$ &\scriptsize\boldmath{$1.36{\pm}1.01$} &\scriptsize$2.50{\pm}1.16$ &\scriptsize$2.64{\pm}4.32$ &\scriptsize$2.17{\pm}1.89$ &- &\scriptsize$2.85{\pm}4.96$ &\scriptsize$2.70{\pm}4.39$ &\scriptsize\tcb{$1.86{\pm}2.12$} &\scriptsize$0.01$\scriptsize${\pm}0.01$ &\scriptsize$0.18$\scriptsize${\pm}0.05$\\

\hline

\end{tabular}}
\end{table*}

\section{Discussion} \label{sec::discussion}

%\subsection{Artificial generation}

In this paper we have shown that training a CNN with sufficiently realistic artificially generated displacement fields, can yield accurate registration results even in real cases. We utilized some randomly generated deformations (single and mixed frequencies) and a more realistic one (respiratory motion). 
We observed that even training with randomly generated deformations in the SPREAD study, the obtained TRE was on par with the B-spline registration (see Table \ref{tb:Test-SPREAD}). Adding more realistic DVFs (respiratory motion) in the DIR-Lab 4DCT study, improved the TRE results from \mbox{$2.70{\pm}4.39$ mm} (``S+M'') to \mbox{$1.86{\pm}2.12$ mm} (``S+M+R'') as can be seen in Table \ref{tb:Test-DIR-Lab-4DCT}. In the case that sufficient realism was not added to the training, for instance in the DIR-Lab-COPDgene study in Table \ref{tb:validation}, the results were sub-optimal. Note that this dataset is challenging for conventional methods also. Anatomical structures in the baseline and follow-up images of this database are quite different and the proposed intensity simulations in Section \ref{sec:method:ag:artificial_deformed_image} did not cover this issue. A solution may be the addition of random intensity occlusions to the deformed images. Another interesting research direction is to learn realistic appearance and deformations from a database.

%
%\rule{0pt}{3ex}Total &\scriptsize$6.62{\pm}3.17^\dagger$ &\scriptsize\tcb{$1.58{\pm}1.29$} &\scriptsize$0.00$\scriptsize${\pm}0.00$ &\scriptsize$0.12$\scriptsize${\pm}0.03$ &\scriptsize$1.60{\pm}1.17^\dagger$ &\scriptsize\boldmath{$1.57{\pm}1.15$} &\scriptsize$1.63{\pm}1.19^\dagger$ &\scriptsize$0.04$\scriptsize${\pm}0.05$ &\scriptsize$0.19$\scriptsize${\pm}0.03$ \\
%\hline
%
%\rule{0pt}{3ex} Total &\scriptsize$8.16{\pm}6.76^\dagger$ &\scriptsize\boldmath{$2.21{\pm}5.86$} &\scriptsize$0.00$\scriptsize${\pm}0.00$ &\scriptsize$0.11$\scriptsize${\pm}0.02$ &\scriptsize\tcb{$2.32{\pm}5.33^\dagger$} &\scriptsize$2.39{\pm}5.64^\dagger$ &\scriptsize$2.65{\pm}5.82^\dagger$ &\scriptsize$0.19$\scriptsize${\pm}0.24$ &\scriptsize$0.24$\scriptsize${\pm}0.06$ \\

%We observed that the more realistic artificial generation can improve the registration quality in some cases, especially in the DIR-Lab-4DCT database (see Table \ref{tb:Test-DIR-Lab-4DCT}).
%However, one of the limitation of the proposed method is registering images with different structures such as images in DIR-Lab-COPDgene database. The main reason is that all proposed intensity simulation in Section \ref{sec:method:ag:artificial_deformed_image} do not cover this issue. One solution might be adding some random intensity occlusions to the deformed images. For conventional metric like mutual information and cross correlation, this database is also not a trivial task.

%\subsection{Large displacements}
One of the major challenges of CNN-based image registration is capturing large DVFs, especially for patch-based methods. Using the whole image as an input might be very time consuming in the training phase. Based on our experiments, the maximum deformation that can be detected in patch with size \mbox{$101 \times 101 \times 101$} by the U-Net-advanced is approximately \mbox{7 mm} (the value of $\theta$ in Table \ref{tb:DVF_freq}). By enlarging the DVFs the Huber loss increased substantially. Please note that the maximum deformation $\theta$ is along each axis so the magnitude of a maximum deformation is $2 \times \sqrt{3} \times \theta$.
Adding the original resolution makes the pipeline slower. However, based on the results in Table \ref{tb:validation} it can be concluded that using two stages (U$^4$-Uadv$^2$, TRE: 1.68$\pm$1.15) can achieve similar results in comparison with three stages (U$^4$-Uadv$^2$-Uadv$^1$, TRE: 1.57$\pm$1.15). Similar to conventional registration, the best method on the first stage is not always the best in combination with others. In Table \ref{tb:validation}, the single U$^4$ is worse than Uadv$^4$ and MV$^4$. Conversely, the combination of U$^4$-Uadv$^2$-Uadv$^1$ obtains the best results. All in all, the differences between the architectures are relatively small and the performance of RegNet is more influenced by the artificial images used in training phase.
%\subsection{Non-isotropic voxel size}
%\subsection{Conventional registration}

The performance of RegNet is very close to the conventional registration. However, B-spline registration is the best in the SPREAD test set (case 13 to 21; $2.21\pm5.86$ vs $2.32\pm5.33$) and the method of \mbox{\cite{berendsen2014registration}} performed better in the DIR-Lab-4DCT database ($1.36\pm1.01$ vs $1.86\pm2.12$). On the contrary, the inference time of CNN approaches are much faster than the conventional methods. Potentially, by increasing the training data and generalizing the artificial generation like sliding motion, the performance of the RegNet can be improved.

%\subsection{Future work} \label{sec:discussion:future_work}

In the current implementation of RegNet, all images are resampled to an isotropic voxel size \mbox{$1.0\times 1.0 \times 1.0$ mm}. If resampling is not intended, it might be possible to simply multiply the output of RegNet by the voxel size. However, this approach is not very accurate because the spatial frequency of different voxel size might not be covered by the training data. A more accurate solution could be to include additional input of the voxel size to the network.

In principle, the proposed network design potentially can be utilized to predict the registration quality. Several methods are suggested by conventional learning using handcrafted features \citep{sokooti2016accuracy, muenzing2012supervised} and a preliminary result by \cite{eppenhof2017supervised}. 

The proposed method can be trained and evaluated on other image modalities like brain MRI images. Potentially, the same network design and artificial generation excluding respiratory motion can be utilized.

The artificial generation can be enhanced if rib segmentation is available. Then, it is possible to incorporate rigid deformation outside of the rib and non-rigid deformations inside the rib. The network potentially can learn the relation between organs and rigidity of the deformations. More realistic and complex simulation like sliding motion of lungs \citep{berendsen2014registration} can also be added to the training images as it had a positive effect for non-learning based methods.

\section{Conclusion}

We proposed a 3D multi-stage CNN framework for chest CT registration. For training the network,
we proposed models to generate artificial DVFs, and intensity models, to easily generate large quantities of paired images with a known spatial relation.
We showed via multiple chest CT databases that this way of artificial training is very effective, with good results on real data. On the public DIR-Lab-4DCT database, we achieved the best results among the CNN approaches.

%
%Several combinations of three proposed network (U-Net, Multi-view and U-Net-advanced) are evaluated in the SPREAD database using several distinctive landmarks. On the DIR-Lab-4DCT database, we achieved the best results among other CNN approaches.

%\section*{Acknowledgments}

%\clearpage

\bibliographystyle{elsarticle-num-names}
\interlinepenalty=10000
\bibliography{RegNet-arXiv}

\begin{thebibliography}{41}
\providecommand{\natexlab}[1]{#1}
\providecommand{\url}[1]{\texttt{#1}}
\providecommand{\urlprefix}{URL }
\expandafter\ifx\csname urlstyle\endcsname\relax
  \providecommand{\doi}[1]{doi:\discretionary{}{}{}#1}\else
  \providecommand{\doi}[1]{doi:\discretionary{}{}{}\begingroup
  \urlstyle{rm}\url{#1}\endgroup}\fi
\providecommand{\bibinfo}[2]{#2}

\bibitem[{Guetter et~al.(2005)Guetter, Xu, Sauer, and
  Hornegger}]{guetter2005learning}
\bibinfo{author}{C.~Guetter}, \bibinfo{author}{C.~Xu},
  \bibinfo{author}{F.~Sauer}, \bibinfo{author}{J.~Hornegger},
  \bibinfo{title}{Learning based non-rigid multi-modal image registration using
  Kullback-Leibler divergence}, in: \bibinfo{booktitle}{International
  Conference on Medical Image Computing and Computer-Assisted Intervention},
  \bibinfo{organization}{Springer}, \bibinfo{pages}{255--262},
  \bibinfo{year}{2005}.

\bibitem[{Jiang et~al.(2008)Jiang, Zheng, Toga, and Tu}]{jiang2008learning}
\bibinfo{author}{J.~Jiang}, \bibinfo{author}{S.~Zheng}, \bibinfo{author}{A.~W.
  Toga}, \bibinfo{author}{Z.~Tu}, \bibinfo{title}{Learning based coarse-to-fine
  image registration}, in: \bibinfo{booktitle}{Computer Vision and Pattern
  Recognition, 2008. CVPR 2008. IEEE Conference on},
  \bibinfo{organization}{IEEE}, \bibinfo{pages}{1--7}, \bibinfo{year}{2008}.

\bibitem[{Hu et~al.(2017)Hu, Wei, Gao, Guo, Wu, and Shen}]{hu2017learning}
\bibinfo{author}{S.~Hu}, \bibinfo{author}{L.~Wei}, \bibinfo{author}{Y.~Gao},
  \bibinfo{author}{Y.~Guo}, \bibinfo{author}{G.~Wu}, \bibinfo{author}{D.~Shen},
  \bibinfo{title}{Learning-based deformable image registration for infant {MR}
  images in the first year of life}, \bibinfo{journal}{Medical Physics}
  \bibinfo{volume}{44}~(\bibinfo{number}{1}) (\bibinfo{year}{2017})
  \bibinfo{pages}{158--170}.

\bibitem[{Muenzing et~al.(2012)Muenzing, van Ginneken, Murphy, and
  Pluim}]{muenzing2012supervised}
\bibinfo{author}{S.~E. Muenzing}, \bibinfo{author}{B.~van Ginneken},
  \bibinfo{author}{K.~Murphy}, \bibinfo{author}{J.~P. Pluim},
  \bibinfo{title}{Supervised quality assessment of medical image registration:
  Application to intra-patient {CT} lung registration},
  \bibinfo{journal}{Medical image analysis}
  \bibinfo{volume}{16}~(\bibinfo{number}{8}) (\bibinfo{year}{2012})
  \bibinfo{pages}{1521--1531}.

\bibitem[{Sokooti et~al.(2016)Sokooti, Saygili, Glocker, Lelieveldt, and
  Staring}]{sokooti2016accuracy}
\bibinfo{author}{H.~Sokooti}, \bibinfo{author}{G.~Saygili},
  \bibinfo{author}{B.~Glocker}, \bibinfo{author}{B.~P. Lelieveldt},
  \bibinfo{author}{M.~Staring}, \bibinfo{title}{Accuracy Estimation for Medical
  Image Registration Using Regression Forests}, in:
  \bibinfo{booktitle}{International Conference on Medical Image Computing and
  Computer-Assisted Intervention}, \bibinfo{organization}{Springer},
  \bibinfo{pages}{107--115}, \bibinfo{year}{2016}.

\bibitem[{Sokooti et~al.(2019)Sokooti, Saygili, Glocker, Lelieveldt, and
  Staring}]{sokooti2019quantitative}
\bibinfo{author}{H.~Sokooti}, \bibinfo{author}{G.~Saygili},
  \bibinfo{author}{B.~Glocker}, \bibinfo{author}{B.~P. Lelieveldt},
  \bibinfo{author}{M.~Staring}, \bibinfo{title}{Quantitative Error Prediction
  of Medical Image Registration using Regression Forests},
  \bibinfo{journal}{Med. Image Anal.} .

\bibitem[{Miao et~al.(2016)Miao, Wang, and Liao}]{miao2016cnn}
\bibinfo{author}{S.~Miao}, \bibinfo{author}{Z.~J. Wang},
  \bibinfo{author}{R.~Liao}, \bibinfo{title}{A {CNN} regression approach for
  real-time 2{D}/3{D} registration}, \bibinfo{journal}{IEEE Transactions on
  Medical Imaging} \bibinfo{volume}{35}~(\bibinfo{number}{5})
  (\bibinfo{year}{2016}) \bibinfo{pages}{1352--1363}.

\bibitem[{Yang et~al.(2016)Yang, Kwitt, and Niethammer}]{yang2016fast}
\bibinfo{author}{X.~Yang}, \bibinfo{author}{R.~Kwitt},
  \bibinfo{author}{M.~Niethammer}, \bibinfo{title}{Fast Predictive Image
  Registration}, in: \bibinfo{booktitle}{International Workshop on Large-Scale
  Annotation of Biomedical Data and Expert Label Synthesis},
  \bibinfo{pages}{48--57}, \bibinfo{year}{2016}.

\bibitem[{Cao et~al.(2017)Cao, Yang, Zhang, Nie, Kim, Wang, and
  Shen}]{cao2017deformable}
\bibinfo{author}{X.~Cao}, \bibinfo{author}{J.~Yang},
  \bibinfo{author}{J.~Zhang}, \bibinfo{author}{D.~Nie},
  \bibinfo{author}{M.~Kim}, \bibinfo{author}{Q.~Wang},
  \bibinfo{author}{D.~Shen}, \bibinfo{title}{Deformable image registration
  based on similarity-steered CNN regression}, in:
  \bibinfo{booktitle}{International Conference on Medical Image Computing and
  Computer-Assisted Intervention}, \bibinfo{organization}{Springer},
  \bibinfo{pages}{300--308}, \bibinfo{year}{2017}.

\bibitem[{Simonovsky et~al.(2016)Simonovsky, Guti{\'e}rrez-Becker, Mateus,
  Navab, and Komodakis}]{simonovsky2016deep}
\bibinfo{author}{M.~Simonovsky}, \bibinfo{author}{B.~Guti{\'e}rrez-Becker},
  \bibinfo{author}{D.~Mateus}, \bibinfo{author}{N.~Navab},
  \bibinfo{author}{N.~Komodakis}, \bibinfo{title}{A deep metric for multimodal
  registration}, in: \bibinfo{booktitle}{International Conference on Medical
  Image Computing and Computer-Assisted Intervention},
  \bibinfo{organization}{Springer}, \bibinfo{pages}{10--18},
  \bibinfo{year}{2016}.

\bibitem[{de~Vos et~al.(2017)de~Vos, Berendsen, Viergever, Staring, and
  I{\v{s}}gum}]{de2017end}
\bibinfo{author}{B.~D. de~Vos}, \bibinfo{author}{F.~F. Berendsen},
  \bibinfo{author}{M.~A. Viergever}, \bibinfo{author}{M.~Staring},
  \bibinfo{author}{I.~I{\v{s}}gum}, \bibinfo{title}{End-to-end unsupervised
  deformable image registration with a convolutional neural network}, in:
  \bibinfo{booktitle}{Deep Learning in Medical Image Analysis and Multimodal
  Learning for Clinical Decision Support}, \bibinfo{publisher}{Springer},
  \bibinfo{pages}{204--212}, \bibinfo{year}{2017}.

\bibitem[{de~Vos et~al.(2019)de~Vos, Berendsen, Viergever, Sokooti, Staring,
  and I{\v{s}}gum}]{de2019deep}
\bibinfo{author}{B.~D. de~Vos}, \bibinfo{author}{F.~F. Berendsen},
  \bibinfo{author}{M.~A. Viergever}, \bibinfo{author}{H.~Sokooti},
  \bibinfo{author}{M.~Staring}, \bibinfo{author}{I.~I{\v{s}}gum},
  \bibinfo{title}{A deep learning framework for unsupervised affine and
  deformable image registration}, \bibinfo{journal}{Med. Image Anal.}
  \bibinfo{volume}{52} (\bibinfo{year}{2019}) \bibinfo{pages}{128--143}.

\bibitem[{Balakrishnan et~al.(2018)Balakrishnan, Zhao, Sabuncu, Guttag, and
  Dalca}]{balakrishnan2018unsupervised}
\bibinfo{author}{G.~Balakrishnan}, \bibinfo{author}{A.~Zhao},
  \bibinfo{author}{M.~R. Sabuncu}, \bibinfo{author}{J.~Guttag},
  \bibinfo{author}{A.~V. Dalca}, \bibinfo{title}{An Unsupervised Learning Model
  for Deformable Medical Image Registration}, \bibinfo{journal}{arXiv preprint
  arXiv:1802.02604} .

\bibitem[{Ferrante et~al.(2018)Ferrante, Oktay, Glocker, and
  Milone}]{ferrante2018adaptability}
\bibinfo{author}{E.~Ferrante}, \bibinfo{author}{O.~Oktay},
  \bibinfo{author}{B.~Glocker}, \bibinfo{author}{D.~H. Milone},
  \bibinfo{title}{On the Adaptability of Unsupervised CNN-Based Deformable
  Image Registration to Unseen Image Domains}, in:
  \bibinfo{booktitle}{International Workshop on Machine Learning in Medical
  Imaging}, \bibinfo{organization}{Springer}, \bibinfo{pages}{294--302},
  \bibinfo{year}{2018}.

\bibitem[{Mahapatra et~al.(2018)Mahapatra, Ge, Sedai, and
  Chakravorty}]{mahapatra2018joint}
\bibinfo{author}{D.~Mahapatra}, \bibinfo{author}{Z.~Ge},
  \bibinfo{author}{S.~Sedai}, \bibinfo{author}{R.~Chakravorty},
  \bibinfo{title}{Joint Registration And Segmentation Of Xray Images Using
  Generative Adversarial Networks}, in: \bibinfo{booktitle}{International
  Workshop on Machine Learning in Medical Imaging},
  \bibinfo{organization}{Springer}, \bibinfo{pages}{73--80},
  \bibinfo{year}{2018}.

\bibitem[{Elmahdy et~al.(2019)Elmahdy, Wolterink, Sokooti, I{\v{s}}gum, and
  Staring}]{elmahdy2019adversarial}
\bibinfo{author}{M.~S. Elmahdy}, \bibinfo{author}{J.~M. Wolterink},
  \bibinfo{author}{H.~Sokooti}, \bibinfo{author}{I.~I{\v{s}}gum},
  \bibinfo{author}{M.~Staring}, \bibinfo{title}{Adversarial optimization for
  joint registration and segmentation in prostate CT radiotherapy},
  \bibinfo{journal}{arXiv preprint arXiv:1906.12223} .

\bibitem[{Sheikhjafari et~al.(2018)Sheikhjafari, Noga, Punithakumar, and
  Ray}]{sheikhjafari2018unsupervised}
\bibinfo{author}{A.~Sheikhjafari}, \bibinfo{author}{M.~Noga},
  \bibinfo{author}{K.~Punithakumar}, \bibinfo{author}{N.~Ray},
  \bibinfo{title}{Unsupervised deformable image registration with fully
  connected generative neural network} .

\bibitem[{Dalca et~al.(2018)Dalca, Balakrishnan, Guttag, and
  Sabuncu}]{dalca2018unsupervised}
\bibinfo{author}{A.~V. Dalca}, \bibinfo{author}{G.~Balakrishnan},
  \bibinfo{author}{J.~Guttag}, \bibinfo{author}{M.~R. Sabuncu},
  \bibinfo{title}{Unsupervised Learning for Fast Probabilistic Diffeomorphic
  Registration}, \bibinfo{journal}{arXiv preprint arXiv:1805.04605} .

\bibitem[{Hu et~al.(2018{\natexlab{a}})Hu, Modat, Gibson, Ghavami, Bonmati,
  Moore, Emberton, Noble, Barratt, and Vercauteren}]{hu2018label}
\bibinfo{author}{Y.~Hu}, \bibinfo{author}{M.~Modat},
  \bibinfo{author}{E.~Gibson}, \bibinfo{author}{N.~Ghavami},
  \bibinfo{author}{E.~Bonmati}, \bibinfo{author}{C.~M. Moore},
  \bibinfo{author}{M.~Emberton}, \bibinfo{author}{J.~A. Noble},
  \bibinfo{author}{D.~C. Barratt}, \bibinfo{author}{T.~Vercauteren},
  \bibinfo{title}{Label-driven weakly-supervised learning for multimodal
  deformarle image registration}, in: \bibinfo{booktitle}{Biomedical Imaging
  (ISBI 2018), 2018 IEEE 15th International Symposium on},
  \bibinfo{organization}{IEEE}, \bibinfo{pages}{1070--1074},
  \bibinfo{year}{2018}{\natexlab{a}}.

\bibitem[{Sokooti et~al.(2017)Sokooti, de~Vos, Berendsen, Lelieveldt,
  I{\v{s}}gum, and Staring}]{sokooti2017nonrigid}
\bibinfo{author}{H.~Sokooti}, \bibinfo{author}{B.~de~Vos},
  \bibinfo{author}{F.~Berendsen}, \bibinfo{author}{B.~P. Lelieveldt},
  \bibinfo{author}{I.~I{\v{s}}gum}, \bibinfo{author}{M.~Staring},
  \bibinfo{title}{Nonrigid Image Registration Using Multi-scale 3D
  Convolutional Neural Networks}, in: \bibinfo{booktitle}{International
  Conference on Medical Image Computing and Computer-Assisted Intervention},
  \bibinfo{organization}{Springer}, \bibinfo{pages}{232--239},
  \bibinfo{year}{2017}.

\bibitem[{Roh{\'e} et~al.(2017)Roh{\'e}, Datar, Heimann, Sermesant, and
  Pennec}]{rohe2017svf}
\bibinfo{author}{M.-M. Roh{\'e}}, \bibinfo{author}{M.~Datar},
  \bibinfo{author}{T.~Heimann}, \bibinfo{author}{M.~Sermesant},
  \bibinfo{author}{X.~Pennec}, \bibinfo{title}{SVF-Net: Learning Deformable
  Image Registration Using Shape Matching}, in:
  \bibinfo{booktitle}{International Conference on Medical Image Computing and
  Computer-Assisted Intervention}, \bibinfo{organization}{Springer},
  \bibinfo{pages}{266--274}, \bibinfo{year}{2017}.

\bibitem[{Fan et~al.(2018)Fan, Cao, Xue, Yap, and Shen}]{fan2018adversarial}
\bibinfo{author}{J.~Fan}, \bibinfo{author}{X.~Cao}, \bibinfo{author}{Z.~Xue},
  \bibinfo{author}{P.-T. Yap}, \bibinfo{author}{D.~Shen},
  \bibinfo{title}{Adversarial Similarity Network for Evaluating Image Alignment
  in Deep Learning Based Registration}, in: \bibinfo{booktitle}{International
  Conference on Medical Image Computing and Computer-Assisted Intervention},
  \bibinfo{organization}{Springer}, \bibinfo{pages}{739--746},
  \bibinfo{year}{2018}.

\bibitem[{Eppenhof et~al.(2018)Eppenhof, Lafarge, Moeskops, Veta, and
  Pluim}]{eppenhof2018deformable}
\bibinfo{author}{K.~A. Eppenhof}, \bibinfo{author}{M.~W. Lafarge},
  \bibinfo{author}{P.~Moeskops}, \bibinfo{author}{M.~Veta},
  \bibinfo{author}{J.~P. Pluim}, \bibinfo{title}{Deformable image registration
  using convolutional neural networks}, in: \bibinfo{booktitle}{Medical Imaging
  2018: Image Processing}, vol. \bibinfo{volume}{10574},
  \bibinfo{organization}{International Society for Optics and Photonics},
  \bibinfo{pages}{105740S}, \bibinfo{year}{2018}.

\bibitem[{Uzunova et~al.(2017)Uzunova, Wilms, Handels, and
  Ehrhardt}]{uzunova2017training}
\bibinfo{author}{H.~Uzunova}, \bibinfo{author}{M.~Wilms},
  \bibinfo{author}{H.~Handels}, \bibinfo{author}{J.~Ehrhardt},
  \bibinfo{title}{Training CNNs for Image Registration from Few Samples with
  Model-based Data Augmentation}, in: \bibinfo{booktitle}{International
  Conference on Medical Image Computing and Computer-Assisted Intervention},
  \bibinfo{organization}{Springer}, \bibinfo{pages}{223--231},
  \bibinfo{year}{2017}.

\bibitem[{Hu et~al.(2018{\natexlab{b}})Hu, Gibson, Ghavami, Bonmati, Moore,
  Emberton, Vercauteren, Noble, and Barratt}]{hu2018adversarial}
\bibinfo{author}{Y.~Hu}, \bibinfo{author}{E.~Gibson},
  \bibinfo{author}{N.~Ghavami}, \bibinfo{author}{E.~Bonmati},
  \bibinfo{author}{C.~M. Moore}, \bibinfo{author}{M.~Emberton},
  \bibinfo{author}{T.~Vercauteren}, \bibinfo{author}{J.~A. Noble},
  \bibinfo{author}{D.~C. Barratt}, \bibinfo{title}{Adversarial Deformation
  Regularization for Training Image Registration Neural Networks},
  \bibinfo{journal}{arXiv preprint arXiv:1805.10665} .

\bibitem[{Ma et~al.(2017)Ma, Wang, Singh, Tamersoy, Chang, Wimmer, and
  Chen}]{ma2017multimodal}
\bibinfo{author}{K.~Ma}, \bibinfo{author}{J.~Wang}, \bibinfo{author}{V.~Singh},
  \bibinfo{author}{B.~Tamersoy}, \bibinfo{author}{Y.-J. Chang},
  \bibinfo{author}{A.~Wimmer}, \bibinfo{author}{T.~Chen},
  \bibinfo{title}{Multimodal Image Registration with Deep Context Reinforcement
  Learning}, in: \bibinfo{booktitle}{International Conference on Medical Image
  Computing and Computer-Assisted Intervention},
  \bibinfo{organization}{Springer}, \bibinfo{pages}{240--248},
  \bibinfo{year}{2017}.

\bibitem[{Krebs et~al.(2017)Krebs, Mansi, Delingette, Zhang, Ghesu, Miao,
  Maier, Ayache, Liao, and Kamen}]{krebs2017robust}
\bibinfo{author}{J.~Krebs}, \bibinfo{author}{T.~Mansi},
  \bibinfo{author}{H.~Delingette}, \bibinfo{author}{L.~Zhang},
  \bibinfo{author}{F.~C. Ghesu}, \bibinfo{author}{S.~Miao},
  \bibinfo{author}{A.~K. Maier}, \bibinfo{author}{N.~Ayache},
  \bibinfo{author}{R.~Liao}, \bibinfo{author}{A.~Kamen}, \bibinfo{title}{Robust
  non-rigid registration through agent-based action learning}, in:
  \bibinfo{booktitle}{International Conference on Medical Image Computing and
  Computer-Assisted Intervention}, \bibinfo{organization}{Springer},
  \bibinfo{pages}{344--352}, \bibinfo{year}{2017}.

\bibitem[{Onieva et~al.(2018)Onieva, Marti-Fuster, de~la Puente, and
  Est{\'e}par}]{onieva2018diffeomorphic}
\bibinfo{author}{J.~O. Onieva}, \bibinfo{author}{B.~Marti-Fuster},
  \bibinfo{author}{M.~P. de~la Puente}, \bibinfo{author}{R.~S.~J. Est{\'e}par},
  \bibinfo{title}{Diffeomorphic Lung Registration Using Deep CNNs and
  Reinforced Learning}, in: \bibinfo{booktitle}{Image Analysis for Moving
  Organ, Breast, and Thoracic Images}, \bibinfo{publisher}{Springer},
  \bibinfo{pages}{284--294}, \bibinfo{year}{2018}.

\bibitem[{Ronneberger et~al.(2015)Ronneberger, Fischer, and
  Brox}]{ronneberger2015u}
\bibinfo{author}{O.~Ronneberger}, \bibinfo{author}{P.~Fischer},
  \bibinfo{author}{T.~Brox}, \bibinfo{title}{U-net: Convolutional networks for
  biomedical image segmentation}, in: \bibinfo{booktitle}{International
  Conference on Medical image computing and computer-assisted intervention},
  \bibinfo{organization}{Springer}, \bibinfo{pages}{234--241},
  \bibinfo{year}{2015}.

\bibitem[{Kamnitsas et~al.(2017)Kamnitsas, Ledig, Newcombe, Simpson, Kane,
  Menon, Rueckert, and Glocker}]{kamnitsas2017efficient}
\bibinfo{author}{K.~Kamnitsas}, \bibinfo{author}{C.~Ledig},
  \bibinfo{author}{V.~F. Newcombe}, \bibinfo{author}{J.~P. Simpson},
  \bibinfo{author}{A.~D. Kane}, \bibinfo{author}{D.~K. Menon},
  \bibinfo{author}{D.~Rueckert}, \bibinfo{author}{B.~Glocker},
  \bibinfo{title}{Efficient multi-scale 3D {CNN} with fully connected {CRF} for
  accurate brain lesion segmentation}, \bibinfo{journal}{Med. Image Anal.}
  \bibinfo{volume}{36} (\bibinfo{year}{2017}) \bibinfo{pages}{61--78}.

\bibitem[{Hub et~al.(2009)Hub, Kessler, and Karger}]{hub2009stochastic}
\bibinfo{author}{M.~Hub}, \bibinfo{author}{M.~L. Kessler},
  \bibinfo{author}{C.~P. Karger}, \bibinfo{title}{A stochastic approach to
  estimate the uncertainty involved in {B}-spline image registration},
  \bibinfo{journal}{IEEE Transactions on Medical Imaging}
  \bibinfo{volume}{28}~(\bibinfo{number}{11}) (\bibinfo{year}{2009})
  \bibinfo{pages}{1708--1716}.

\bibitem[{Staring et~al.(2014)Staring, Bakker, Stolk, Shamonin, Reiber, and
  Stoel}]{staring2014towards}
\bibinfo{author}{M.~Staring}, \bibinfo{author}{M.~Bakker},
  \bibinfo{author}{J.~Stolk}, \bibinfo{author}{D.~Shamonin},
  \bibinfo{author}{J.~Reiber}, \bibinfo{author}{B.~Stoel},
  \bibinfo{title}{Towards local progression estimation of pulmonary emphysema
  using {CT}}, \bibinfo{journal}{Medical Physics}
  \bibinfo{volume}{41}~(\bibinfo{number}{2}) (\bibinfo{year}{2014})
  \bibinfo{pages}{021905}.

\bibitem[{Rueckert et~al.(1999)Rueckert, Sonoda, Hayes, Hill, Leach, and
  Hawkes}]{rueckert1999nonrigid}
\bibinfo{author}{D.~Rueckert}, \bibinfo{author}{L.~I. Sonoda},
  \bibinfo{author}{C.~Hayes}, \bibinfo{author}{D.~L. Hill},
  \bibinfo{author}{M.~O. Leach}, \bibinfo{author}{D.~J. Hawkes},
  \bibinfo{title}{Nonrigid registration using free-form deformations:
  application to breast MR images}, \bibinfo{journal}{IEEE transactions on
  medical imaging} \bibinfo{volume}{18}~(\bibinfo{number}{8})
  (\bibinfo{year}{1999}) \bibinfo{pages}{712--721}.

\bibitem[{Stolk et~al.(2007)Stolk, Putter, Bakker, Shaker, Parr, Piitulainen,
  Russi, Grebski, Dirksen, Stockley, Reiber, and Stoel}]{stolk2007progression}
\bibinfo{author}{J.~Stolk}, \bibinfo{author}{H.~Putter}, \bibinfo{author}{E.~M.
  Bakker}, \bibinfo{author}{S.~B. Shaker}, \bibinfo{author}{D.~G. Parr},
  \bibinfo{author}{E.~Piitulainen}, \bibinfo{author}{E.~W. Russi},
  \bibinfo{author}{E.~Grebski}, \bibinfo{author}{A.~Dirksen},
  \bibinfo{author}{R.~A. Stockley}, \bibinfo{author}{J.~H.~C. Reiber},
  \bibinfo{author}{B.~C. Stoel}, \bibinfo{title}{Progression parameters for
  emphysema: a clinical investigation}, \bibinfo{journal}{Respiratory medicine}
  \bibinfo{volume}{101}~(\bibinfo{number}{9}) (\bibinfo{year}{2007})
  \bibinfo{pages}{1924--1930}.

\bibitem[{Castillo et~al.(2009)Castillo, Castillo, Guerra, Johnson, McPhail,
  Garg, and Guerrero}]{castillo2009framework}
\bibinfo{author}{R.~Castillo}, \bibinfo{author}{E.~Castillo},
  \bibinfo{author}{R.~Guerra}, \bibinfo{author}{V.~E. Johnson},
  \bibinfo{author}{T.~McPhail}, \bibinfo{author}{A.~K. Garg},
  \bibinfo{author}{T.~Guerrero}, \bibinfo{title}{A framework for evaluation of
  deformable image registration spatial accuracy using large landmark point
  sets}, \bibinfo{journal}{Physics in medicine and biology}
  \bibinfo{volume}{54}~(\bibinfo{number}{7}) (\bibinfo{year}{2009})
  \bibinfo{pages}{1849}.

\bibitem[{Castillo et~al.(2013)Castillo, Castillo, Fuentes, Ahmad, Wood,
  Ludwig, and Guerrero}]{castillo2013reference}
\bibinfo{author}{R.~Castillo}, \bibinfo{author}{E.~Castillo},
  \bibinfo{author}{D.~Fuentes}, \bibinfo{author}{M.~Ahmad},
  \bibinfo{author}{A.~M. Wood}, \bibinfo{author}{M.~S. Ludwig},
  \bibinfo{author}{T.~Guerrero}, \bibinfo{title}{A reference dataset for
  deformable image registration spatial accuracy evaluation using the COPDgene
  study archive}, \bibinfo{journal}{Physics in Medicine \& Biology}
  \bibinfo{volume}{58}~(\bibinfo{number}{9}) (\bibinfo{year}{2013})
  \bibinfo{pages}{2861}.

\bibitem[{Murphy et~al.(2011)Murphy, van Ginneken, Klein, Staring, de~Hoop,
  Viergever, and Pluim}]{murphy2011semi}
\bibinfo{author}{K.~Murphy}, \bibinfo{author}{B.~van Ginneken},
  \bibinfo{author}{S.~Klein}, \bibinfo{author}{M.~Staring},
  \bibinfo{author}{B.~J. de~Hoop}, \bibinfo{author}{M.~A. Viergever},
  \bibinfo{author}{J.~P. Pluim}, \bibinfo{title}{Semi-automatic construction of
  reference standards for evaluation of image registration},
  \bibinfo{journal}{Med. Image Anal.}
  \bibinfo{volume}{15}~(\bibinfo{number}{1}) (\bibinfo{year}{2011})
  \bibinfo{pages}{71--84}.

\bibitem[{Berendsen et~al.(2014)Berendsen, Kotte, Viergever, and
  Pluim}]{berendsen2014registration}
\bibinfo{author}{F.~F. Berendsen}, \bibinfo{author}{A.~N. Kotte},
  \bibinfo{author}{M.~A. Viergever}, \bibinfo{author}{J.~P. Pluim},
  \bibinfo{title}{Registration of organs with sliding interfaces and changing
  topologies}, in: \bibinfo{booktitle}{Medical Imaging 2014: Image Processing},
  vol. \bibinfo{volume}{9034}, \bibinfo{organization}{International Society for
  Optics and Photonics}, \bibinfo{pages}{90340E}, \bibinfo{year}{2014}.

\bibitem[{Eppenhof and Pluim(2018)}]{eppenhof2018pulmonary}
\bibinfo{author}{K.~A. Eppenhof}, \bibinfo{author}{J.~P. Pluim},
  \bibinfo{title}{Pulmonary CT Registration through Supervised Learning with
  Convolutional Neural Networks}, \bibinfo{journal}{IEEE transactions on
  Medical Imaging} .

\bibitem[{Sentker et~al.(2018)Sentker, Madesta, and Werner}]{sentker2018gdl}
\bibinfo{author}{T.~Sentker}, \bibinfo{author}{F.~Madesta},
  \bibinfo{author}{R.~Werner}, \bibinfo{title}{GDL-FIRE 4D: Deep Learning-Based
  Fast 4D CT Image Registration}, in: \bibinfo{booktitle}{International
  Conference on Medical Image Computing and Computer-Assisted Intervention},
  \bibinfo{organization}{Springer}, \bibinfo{pages}{765--773},
  \bibinfo{year}{2018}.

\bibitem[{Eppenhof and Pluim(2017)}]{eppenhof2017supervised}
\bibinfo{author}{K.~A. Eppenhof}, \bibinfo{author}{J.~P. Pluim},
  \bibinfo{title}{Supervised local error estimation for nonlinear image
  registration using convolutional neural networks}, in:
  \bibinfo{booktitle}{SPIE Medical Imaging},
  \bibinfo{organization}{International Society for Optics and Photonics},
  \bibinfo{pages}{101331U--101331U}, \bibinfo{year}{2017}.

\end{thebibliography}

\end{document}